\documentclass[aip,,reprint,twocolumn,a4paper]{revtex4-1}

\usepackage{graphicx}
\usepackage{amssymb,amsfonts,amsmath}

\usepackage{subfig}
\usepackage{afterpage}

\usepackage[sort&compress]{natbib}

\begin{document}
\title{\LARGE Identifying States of a Financial Market}

\affiliation{Center of Polymer Studies, Boston University, USA}
\affiliation{Faculty of Physics, University of Duisburg-Essen,  Germany}
\affiliation{Department of Applied Physics, Graduate School of Engineering, The University of \mbox{Tokyo, Japan}}
\affiliation{Instituto de Ciencias F\'isicas, Universidad Nacional Aut\'onoma de M\'exico and Centro Internacional de Ciencias, Cuernavaca, Mexico}

\author{Michael C. M\"unnix$^{1,2}$}
\author{Takashi Shimada$^{1,3}$}
\author{Rudi Sch\"afer$^{2}$}
\author{Francois Leyvraz$^{4}$}
\author{Thomas H. Seligman$^{4}$}
\author{Thomas Guhr$^{2}$}
\author{H. Eugene Stanley$^{1}$}

\date{April 2011}

\begin{abstract}

The understanding of complex systems has become a central issue because complex systems exist in a wide range of scientific disciplines.
Time series are typical experimental results we have about complex systems. In the analysis of such time series, stationary situations have been extensively studied and correlations have been found to be a very powerful tool.
Yet most natural processes are non-stationary. In particular, in times of crisis, accident or trouble, stationarity is lost. As examples we may think of financial markets, biological systems, reactors (both chemical and nuclear) or the weather. In non-stationary situations analysis becomes very difficult and noise is a severe problem. Following a natural urge to search for order in the system, we endeavor to define states through which systems pass and in which they remain for short times. 
Success in this respect would allow to get a better understanding of the system and might even lead to methods for controlling the system in more efficient ways.

We here concentrate on financial markets because of the easy access we have to good data, because of our previous experience and last but not least because of the strong non-stationary effects recently seen. We analyze the S\&P 500 stocks in the 19-year period 1992-2010. Here, we propose such an above mentioned definition of state for a financial market and use it to identify points of drastic change in the correlation structure. These points are mapped to occurrences of financial crises. We find that a wide variety of characteristic correlation structure patterns exist in the observation time window, and that these characteristic correlation structure patterns can be classified into several typical ``market states''. Using this classification we recognize transitions between different market states. A  similarity measure we develop thus affords means of understanding changes in states and of recognizing developments not previously seen.
\end{abstract}

\keywords{Non-stationarity, Market similarity, Market states}

\maketitle

  The effort to understand the dynamics in financial markets is
  attracting scientists from many fields
  \cite{b_voit01,mantegna07,borghesi07,cooley01,pelletier05,xu06,king90,lee-s06}.
  Statistical dependencies between stocks are of particular
  interest, because they play a major role in the estimation of
  financial risk \cite{b_bouchaud00}. Since the market itself is subject to continuous change, 
  the statistical dependencies also change in
  time. This non-stationary behavior makes an analysis very difficult \cite{eckmann87,casdagli97}.
  Changes in supply and demand can even lead to a two phase behavior of the market \cite{plerou03b}. Here, we use the correlation matrix to identify and classify the market state. In particular we ask, how similar is the present market state, compared to previous states? To calculate this \emph{similarity} we measure temporal changes in the statistical dependence between stock returns.
  
For stationary systems described by a (generally large) number $K$ of time series, the Pearson correlation coefficient is extremely useful. It is defined as%
\begin{equation}
C_{ij} \equiv \frac{\left\langle r_{i}r_{j}\right\rangle -\left\langle
  r_{i}\right\rangle\left\langle
  r_{j}\right\rangle}{\sigma_{i}\sigma_{j}} \ .
\label{eq:nat-pearson}
\end{equation}

\enlargethispage{\baselineskip}

Here the $r_{i}$ and $r_{j}$ represent the time series of which the averages $\langle\ldots\rangle$ are taken over a given time horizon T. $\sigma_i$ and $\sigma_j$ are their respective standard deviations. When calculating the
correlation coefficients of $K$ stocks, we obtain the $K\times K$ correlation matrix $\mathbf{C}$, which gives an insight into the statistical interdependencies of the time series under study.

It is necessary to consider data over large time horizons $T$ so as to obtain reliable statistics. This leads to a fundamental problem that arises in the case of \emph{non-stationary} systems: To extract useful information from empirical data we seek a correlation matrix from very recent data, in order to provide a good description of current correlation structure. This is because correlations change dynamically due to the non-stationarity of the process, making it very difficult to estimate them precisely \cite{rosenow03,tastan06,drozdz01,schaefer08}. However, if the length $T$ of the time series is short, the correlation matrices $\mathbf{C}$ are noisy. On the other hand, to keep the estimation error low, $T$ can be increased, but this leads to a correlation matrix that generally does not describe the present state very well. Various noise reduction techniques provide methods to conquer noise \cite{laloux99,plerou03,guhr03,schaferj05,ledoit03}.

\begin{figure}[tb]
\centering
\subfloat[daily data]{\includegraphics[width=0.46\textwidth]{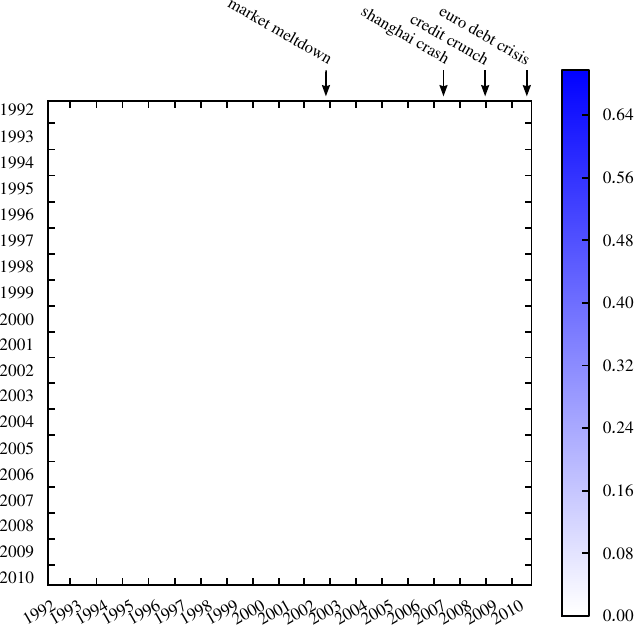}\label{img:nat-marketsim}}
\quad
\subfloat[intraday data]{\includegraphics[width=0.46\textwidth]{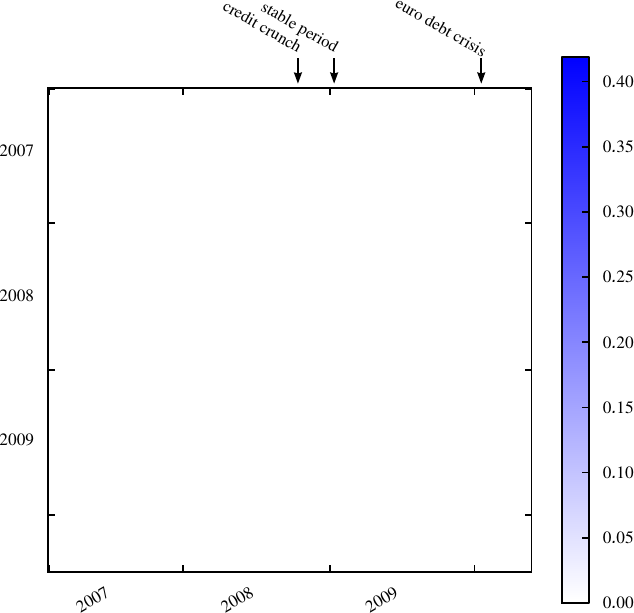}\label{img:nat-marketsimintraday}}
\caption{Financial crisis are accompanied by drastic changes in the
  correlation structure, indicated by blue shaded areas. The market
  similarity $\zeta$ in panel (a) is based on daily data. Panel (b) is a
  more detailed study of the 2007--2010 period, including the ``credit
  crunch'' and the initial impact of the european debt crisis. The area of panel (b) is a magnification of the lower right
  square in panel (a).}
\end{figure}

In several non-stationary systems, it is possible to obtain a large number of correlated data over time. Such systems include, but are not restricted to, financial markets (which show  non-stationary behavior due to crises), biological or medical time series (such as EEG), chemical and nuclear reactors (non-stationary behavior includes, in particular, accidents) or weather data. In the following, we  only consider the financial markets, since we have studied extensively some very high quality data of this system, the non-stationary features of which have been quite striking in  the last years. We propose a definition of  a state which is appropriate for such systems and suggest a method of analysis which allows for a classification of possible behaviors of the system. When $T/K<1$, which is the case we are interested in, the correlation matrix becomes singular. However, one can still make significant statistical statements, e.g., for the average correlation level whose estimation error decreases as $1/K$. In the following, we focus on  correlation matrices  $\mathbf C(t_{1})$ and $\mathbf C(t_{2})$ at different times $t_{1}$ and $t_{2}$ measured over a \emph{short} time horizon. These have therefore a pronounced random element. We take these objects as the fundamental states of our system. We now propose, as a central element, to introduce the following concept of distance between two states. We define

\begin{equation}
\zeta(t_{1},t_{2})\equiv \left\langle\left|C_{ij}(t_{1})-C_{ij}(t_{2})\right|\right\rangle_{ij}
\label{eq:nat-simmeasure}
\end{equation}
to quantify the difference of the correlation structure for two points in time, where $| \ldots |$ denotes the absolute value and $\langle\ldots\rangle_{ij}$ denotes the average over all components. Note that in this case, the random component that is unavoidable in the definition of the states of the system is strongly suppressed by the average over $K^2\gg 1$ numbers.

To apply the above general statements to a specific example, we analyze
two datasets: (i) we calculate $\zeta(t_1,t_2)$ based on the daily
returns of those S\&P 500 stocks that remained part of the S\&P during
the 19-year period 1992--2010, and (ii) we study the four-year period
2007--2010 in more detail based on intraday data from the NYSE~TAQ
database. Since the noise increases for very high-frequency data
\cite{epps79,muennix09b,muennix10a}, we extract one-hour returns for
dataset (ii). For one-hour returns, we consider this market microstructure noise as reasonably weak.

However, sudden changes in drift and volatility are present on all time scales.
They can result in erroneous
correlation estimates. To address this problem, we employ
a local normalization \cite{schaefer10b} of the return time series in dataset (i).
The results of dataset (i) are presented in Fig.~\ref{img:nat-marketsim}.
In this figure, each point is calculated on correlation matrices over the
previous two months.
This new representation gives a complete overview about structural changes of this financial market of the past 19 years in a single figure.
It allows to compare the similarity of the market
states at different times.  To make this procedure concrete, consider
the following example. Pick a point on the diagonal of
Fig.~\ref{img:nat-marketsim} and designate it as ``now''. From this
point the similarity to previous times can be found on the vertical line
above this point, or the horizontal line to the left of this point.
Light shading denotes similar market states and dark shading denotes
dissimilar states. We can furthermore identify times of financial crises
with dark shaded areas. This indicates that the correlation structure
completely changes during a crisis. There are also similarities between
crises, as between the ``credit crunch'' that induced the 2008--2009 financial crisis and the ``market
meltdown'', the burst of the dot-com bubble in 2002. A further example is the overall rise in correlation
level in the beginning of 2007. This event can be mapped to drastic
events on the Shanghai stock exchange \cite{bbw07}.


\begin{figure*}[p]
\subfloat[state 1]{\includegraphics[width=0.23\textwidth]{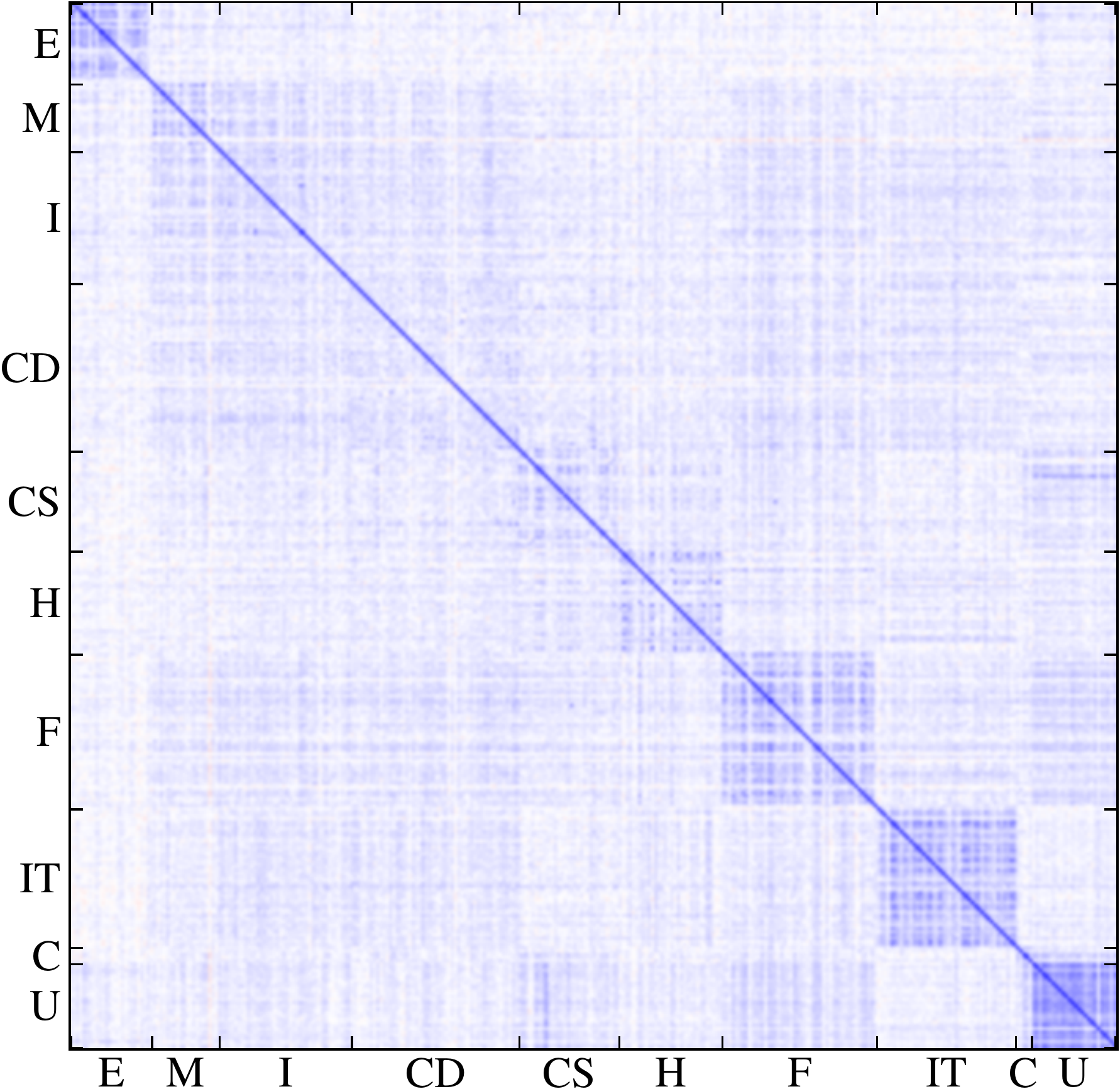}}
\quad
\subfloat[state 2]{\includegraphics[width=0.23\textwidth]{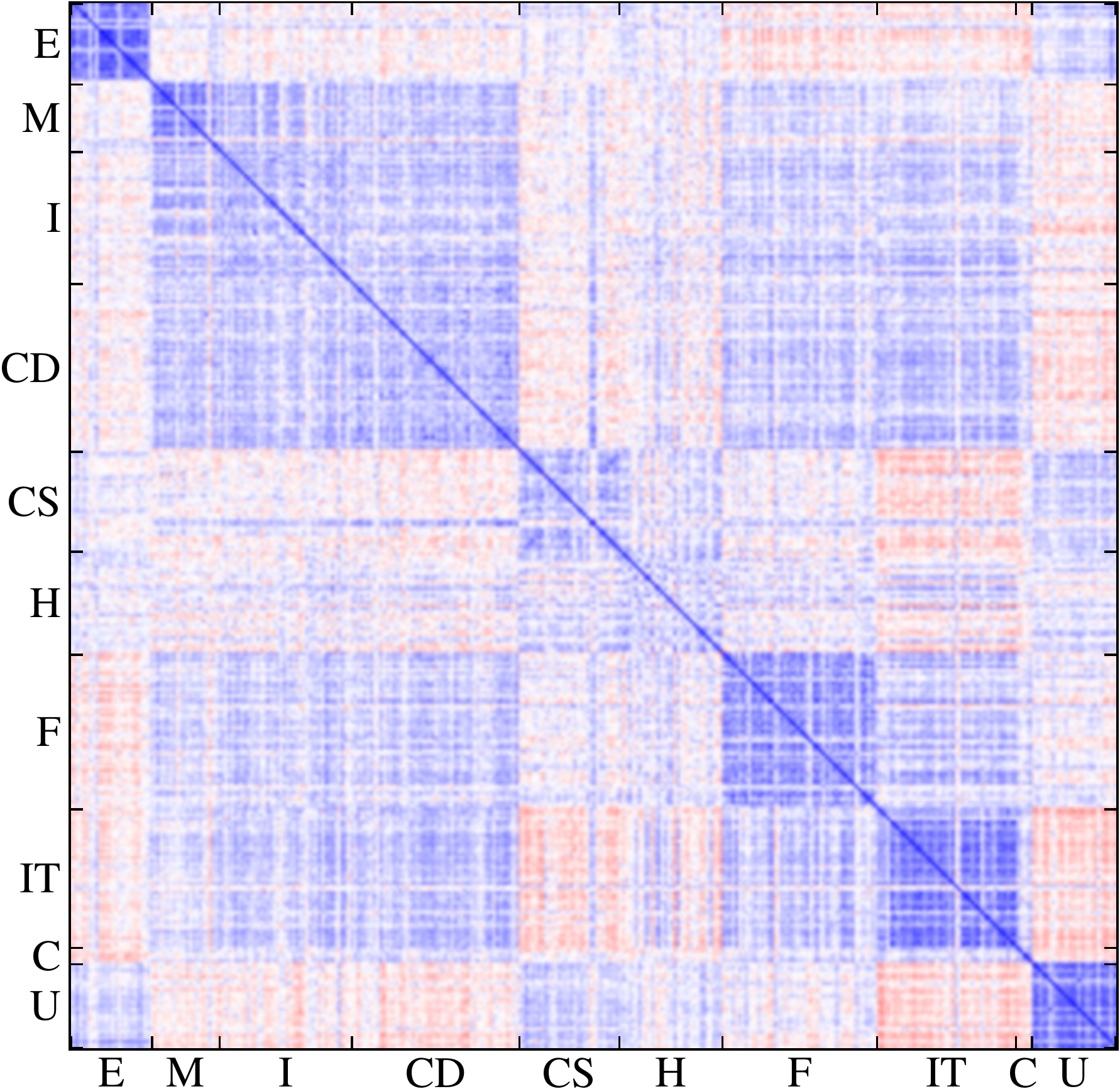}}
\quad
\subfloat[state 3]{\includegraphics[width=0.23\textwidth]{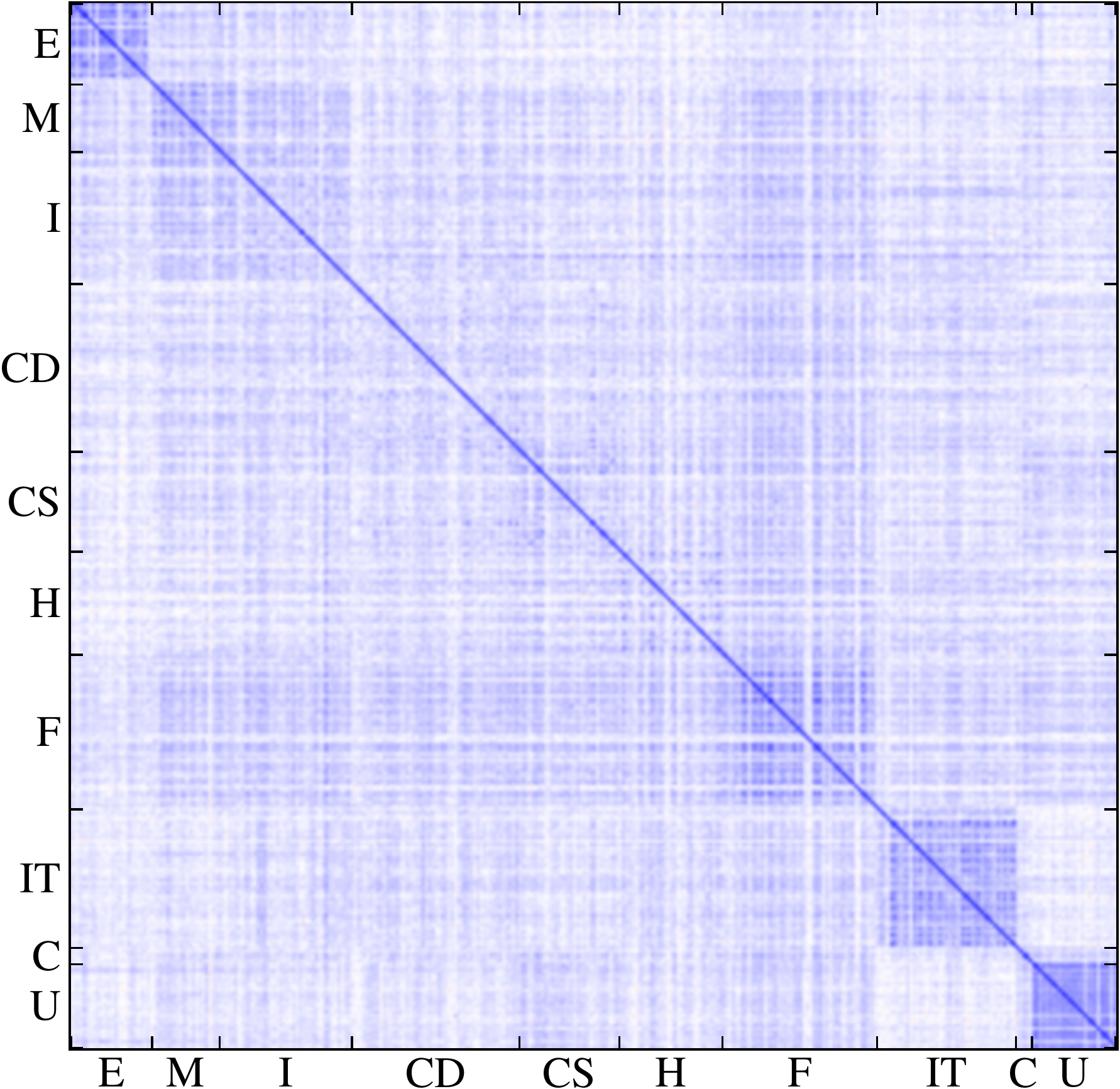}}
\quad
\subfloat[state 4]{\includegraphics[width=0.23\textwidth]{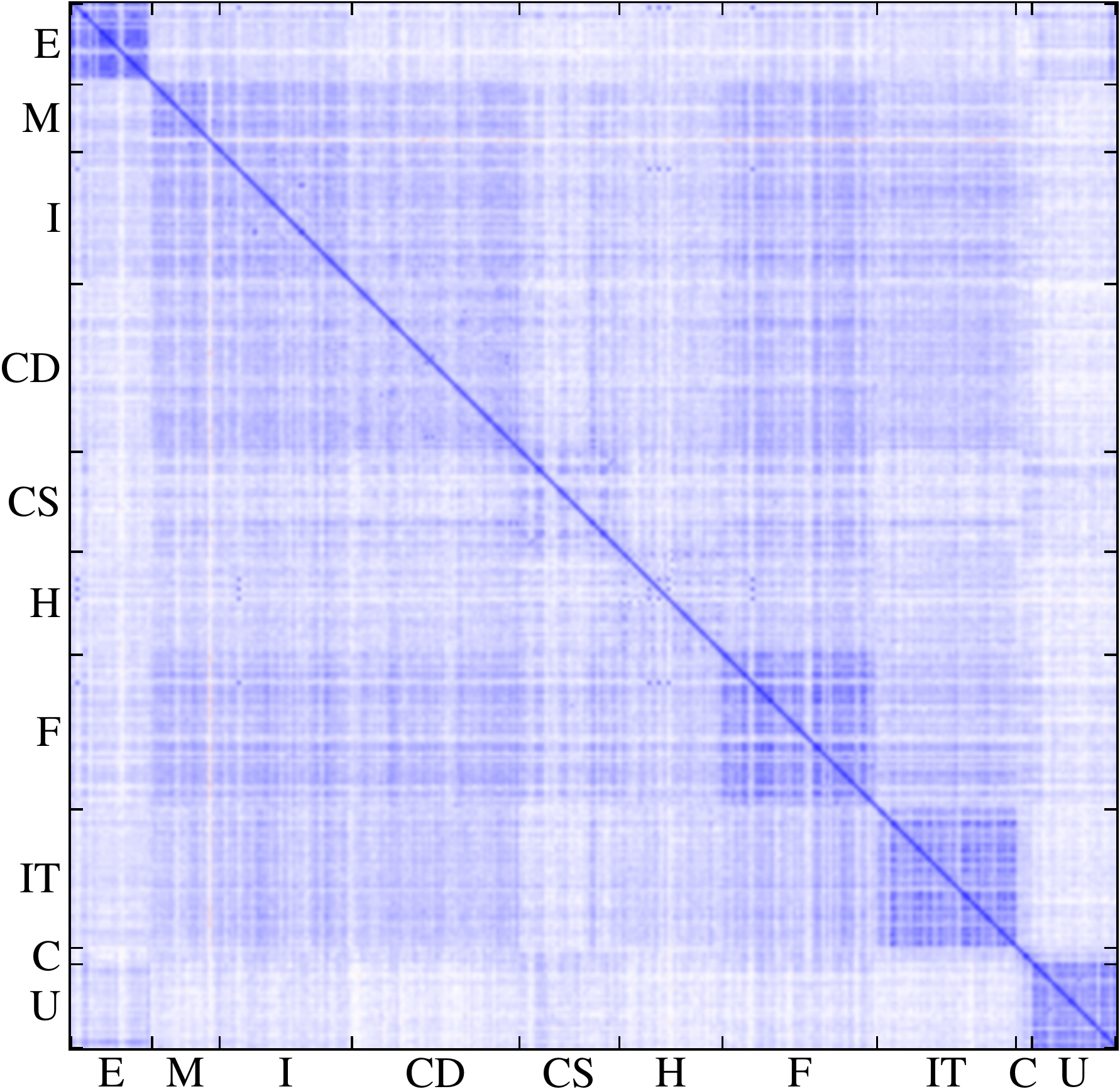}}
\\
\subfloat[state 5]{\includegraphics[width=0.23\textwidth]{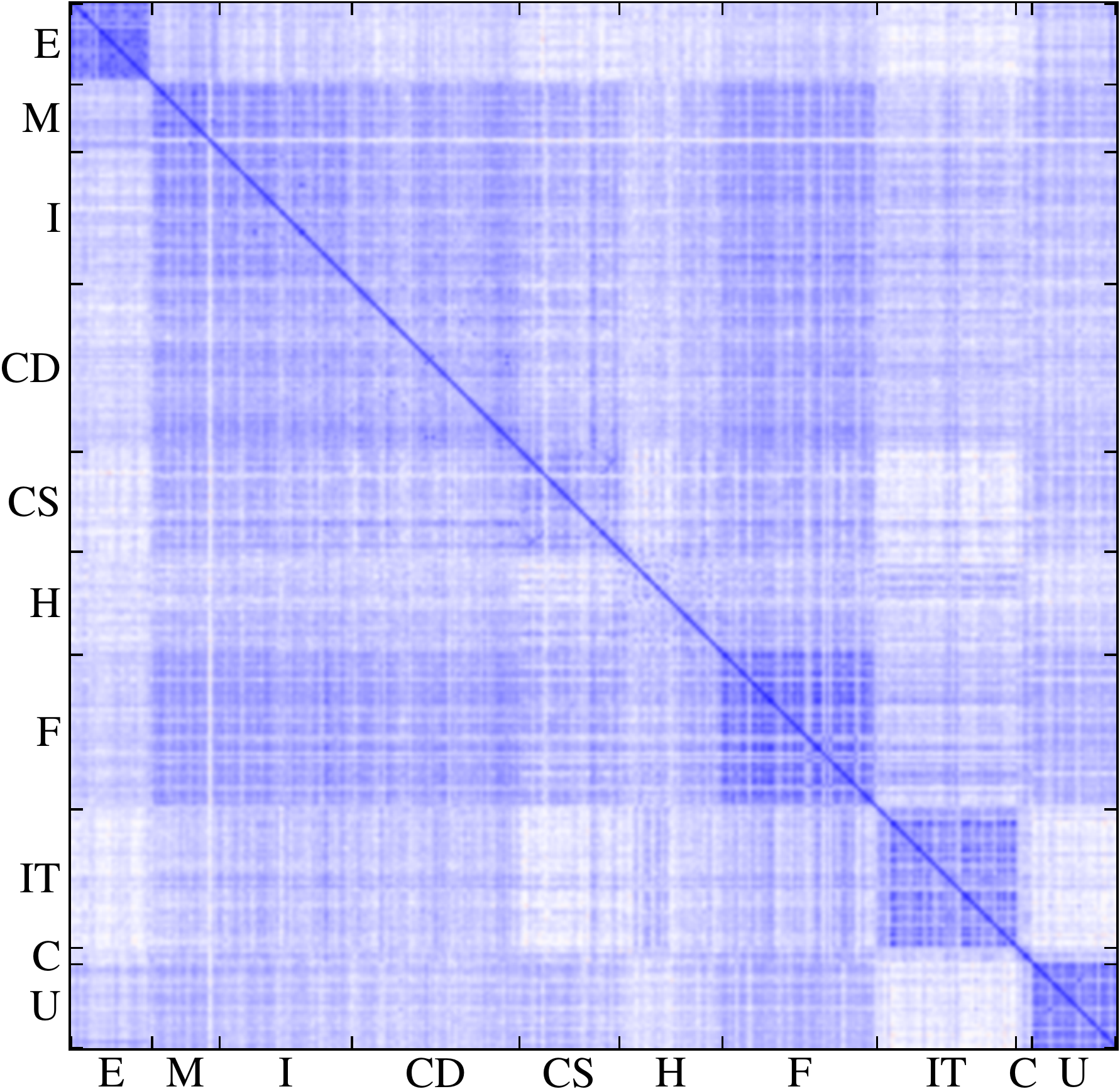}}
\quad
\subfloat[state 6]{\includegraphics[width=0.23\textwidth]{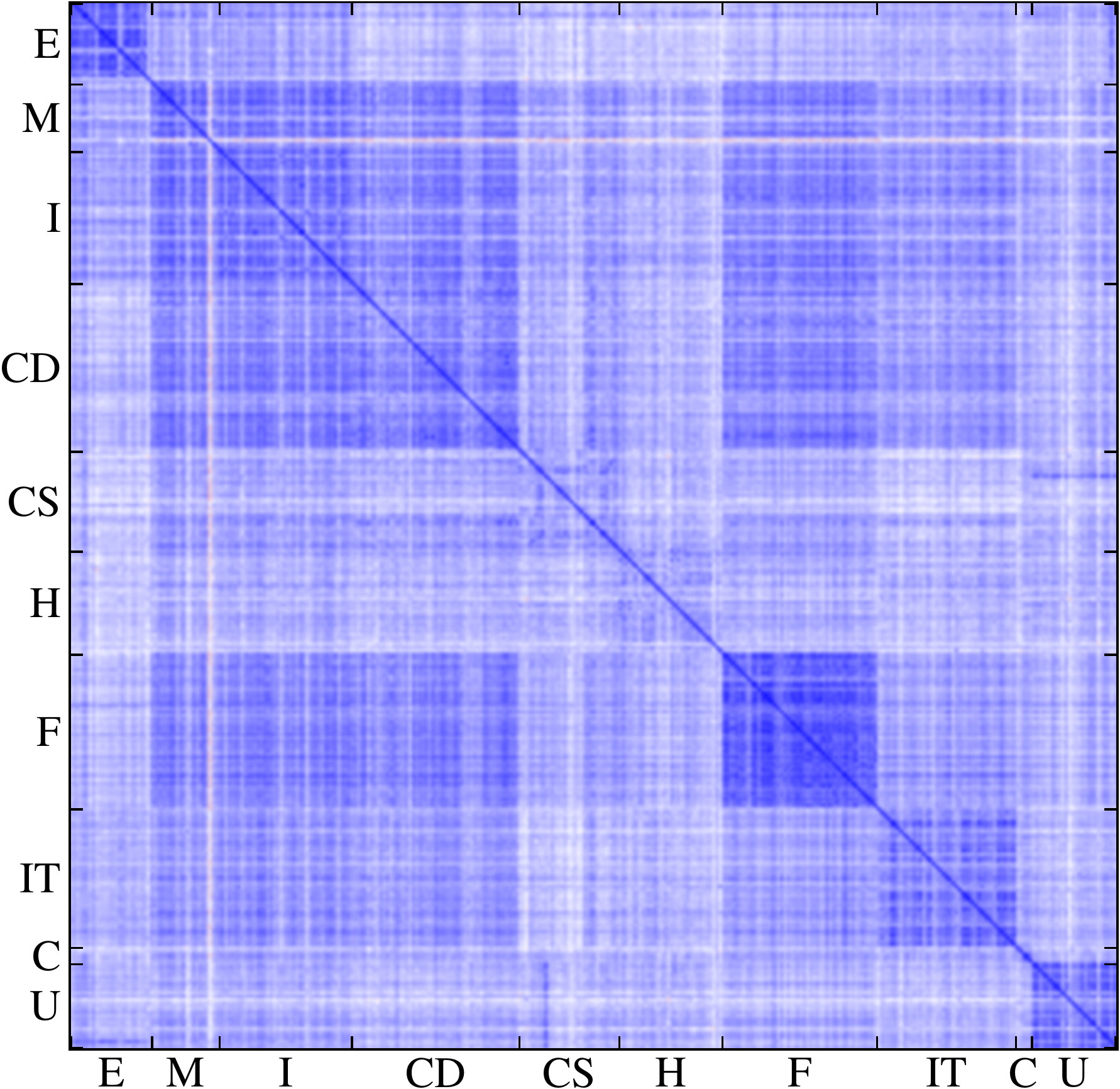}}
\quad
\subfloat[state 7]{\includegraphics[width=0.23\textwidth]{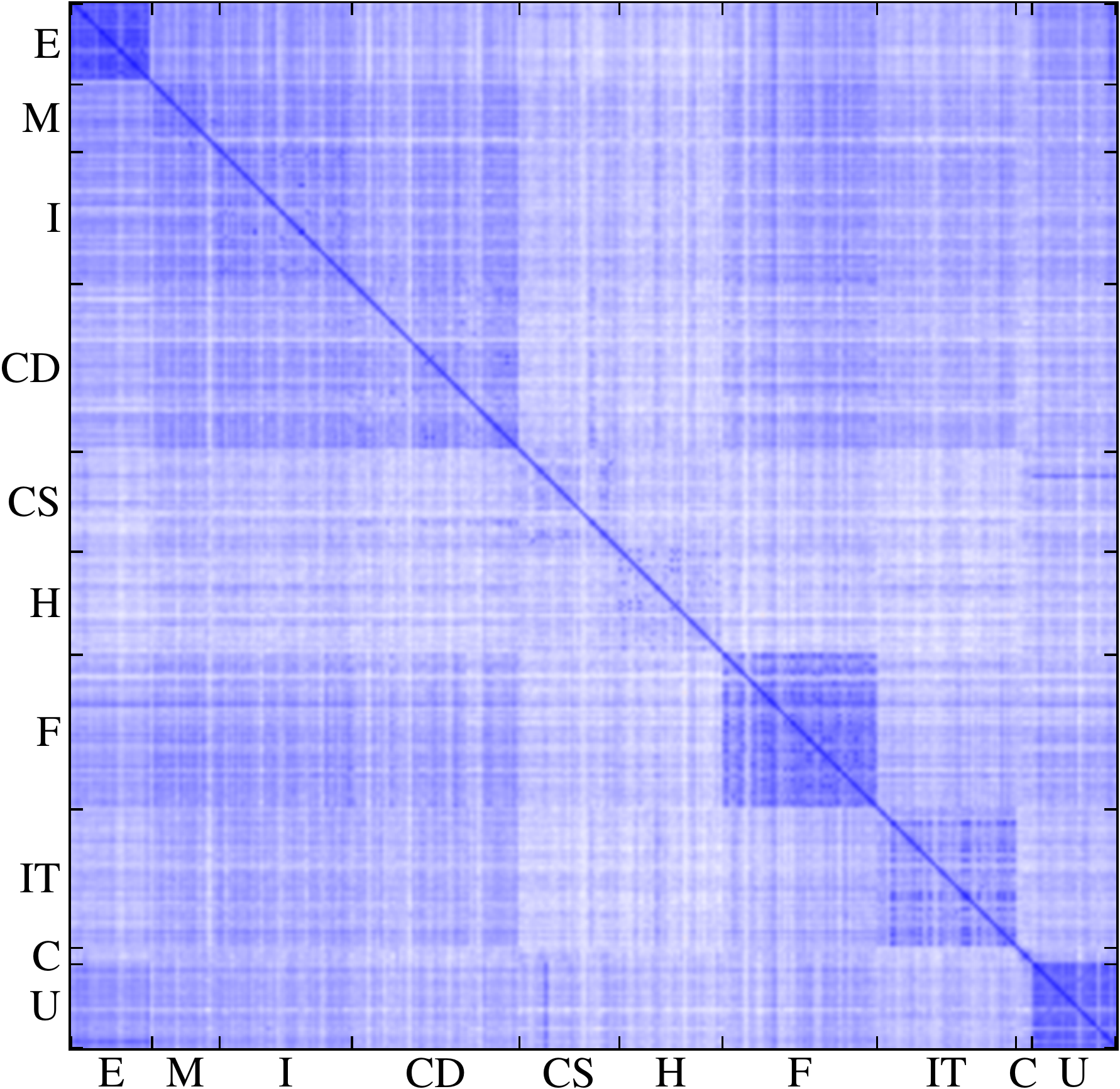}}
\quad
\subfloat[state 8]{\includegraphics[width=0.23\textwidth]{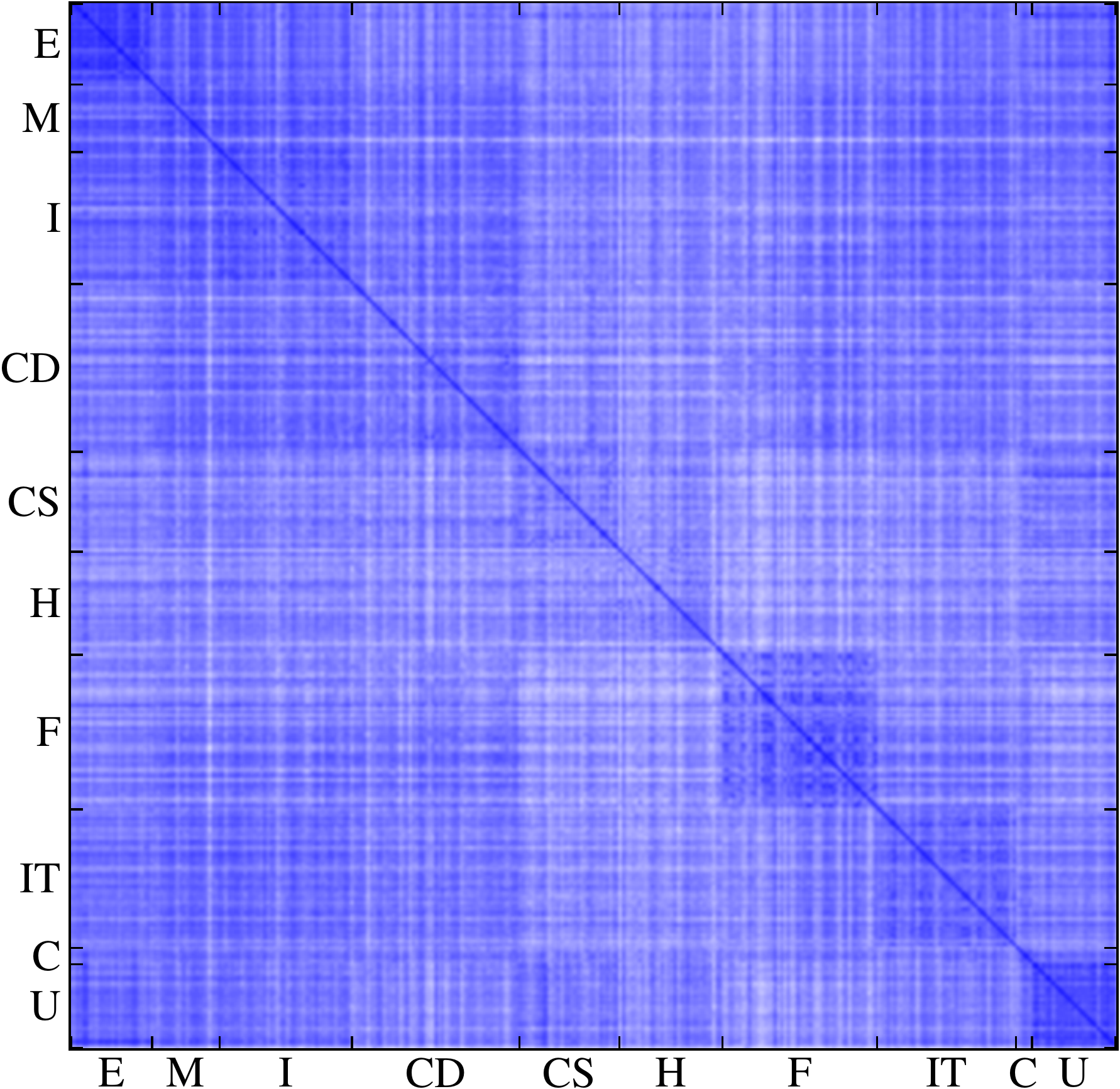}}
\\
\subfloat[Clustering tree]{\includegraphics[width=0.26\textwidth]{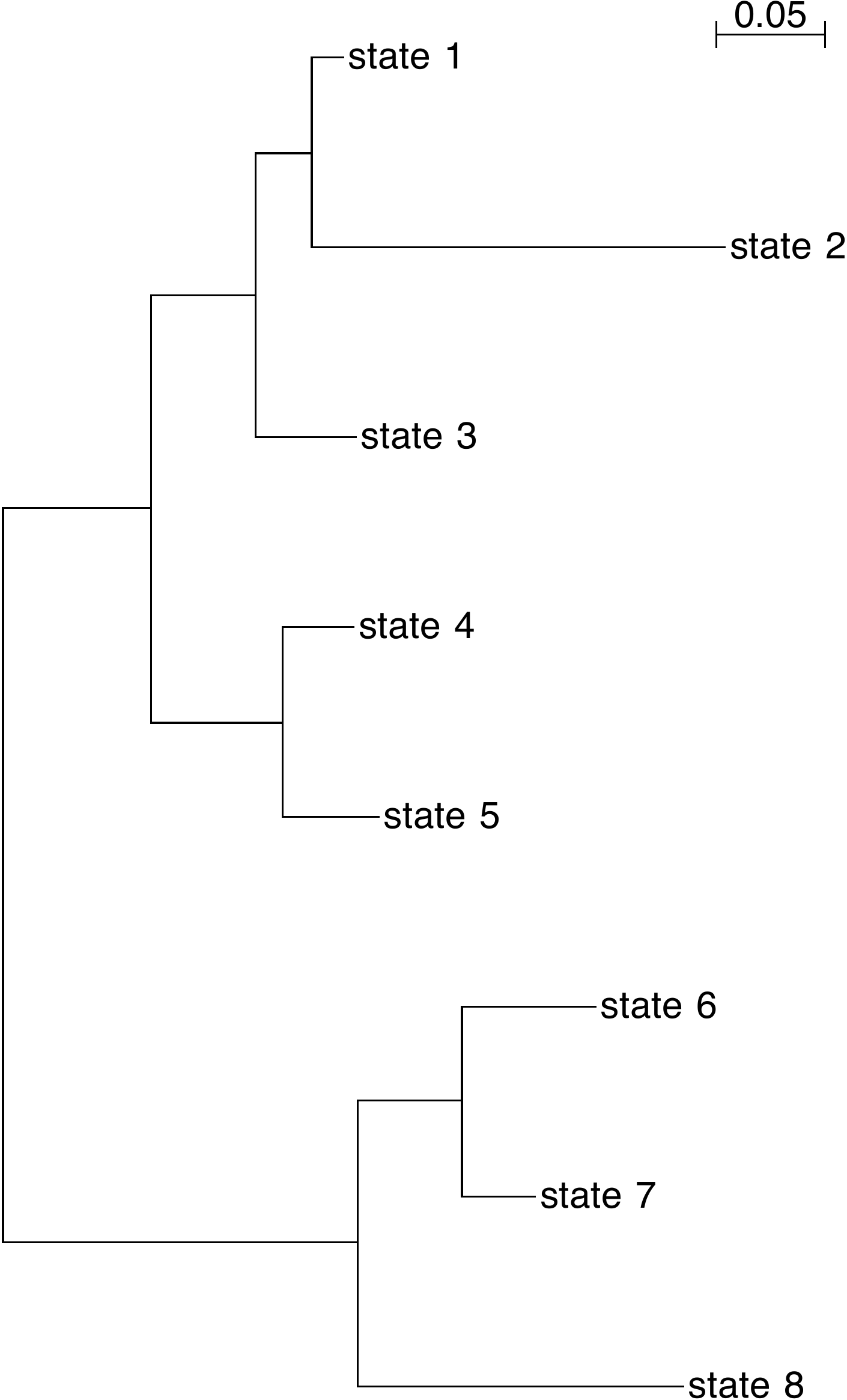}}
\qquad\qquad\qquad\qquad
\subfloat[Overall average
  correlation]{\includegraphics[width=0.45\textwidth]{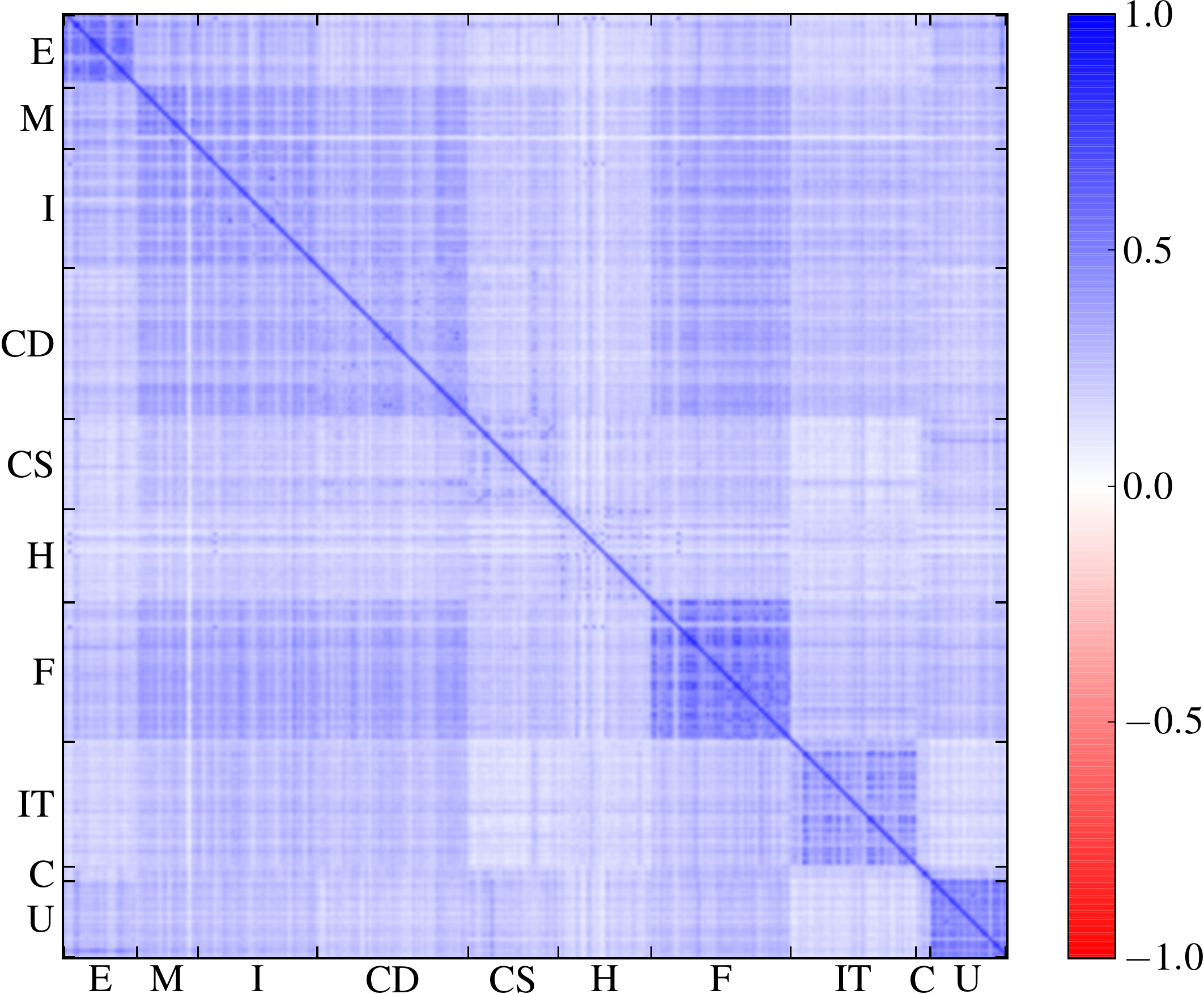}}   
\caption{The correlation between different industry branches as well as
  the intra-branch correlation characterize the different market states
  (a-h). The inter-branch correlation is represented by the off-diagonal
  blocks, and the intra-branch correlation is represented by the blocks
  in the diagonal.  Legend: E: Energy, M: Materials, I: Industrials, CD:
  Consumer Discretionary, CS: Consumer Staples, H: Health Care, F:
  Financials, IT: Information Technology, C: Communication, U:
  Utilities. (i) Similarity tree structure of the 8 market states. (j)
  Illustration of the overall average correlation matrix.}
\label{img:nat-typicalstates}
\end{figure*}


\begin{figure*}[t]
\includegraphics[width=0.95\textwidth]{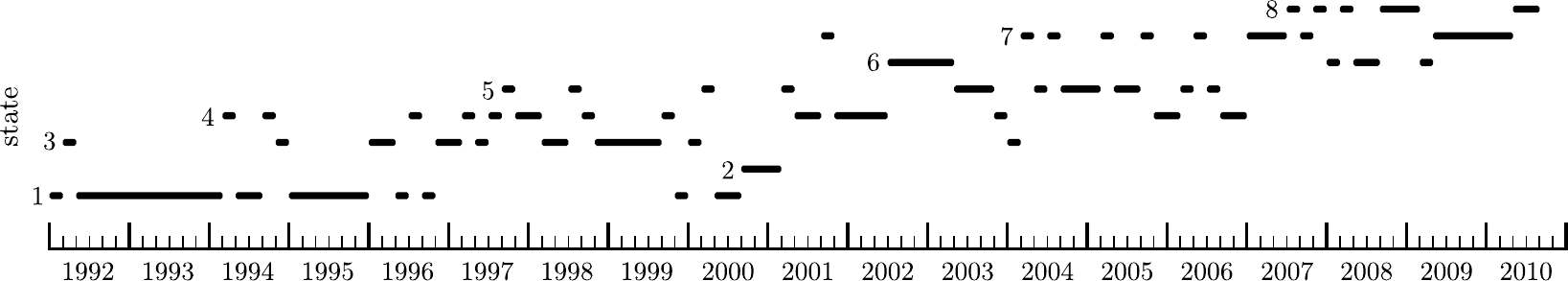}
\caption{Temporal evolution of the market state. The horizontal axis
  represents the observation time and the vertical axis denotes the
  market state obtained from top-down clustering.  The market state
  sometimes remains in the same state for a long time, and sometimes for
  a short time in the same state. It also can return to a state that it
  has previously visited.  Some states (e.g., state 1 and state 2)
  appear to cluster in time, while other states appear more sparsely and
  intermittently in time (e.g., state 4).}
\label{img:nat-stateevolution}
\end{figure*}
\newpage

Using dataset (ii) we are able to obtain a more detailed insight into
recent market changes, as shown in
Fig.~\ref{img:nat-marketsimintraday}. This area is represented by the
lower right square in Fig.~\ref{img:nat-marketsim}. Using intraday data
we calculate the correlation matrices on shorter time scales. We choose a time horizon of one week, which, because it provides
insight into changes in the correlation structure on a much finer time
scale, enables us to identify a short sub-period within the 2008--2009
crisis (in the beginning of 2009) during which the market temporarily
stabilizes before it returns to the crisis state. While the correlation
structure during the crisis displays an overall high correlation level,
the correlation structure of the stable period is similar to the period before the crisis,
one of the typical states in a calm period, which is identified from
daily data in dataset (i). This phenomenon might be related to the market's
reaction to news about the progress in rescuing the American
International Group (A.I.G.) \cite{nytimes08}.  The
correlation structure of this stable period can be found in the
\emph{Supplementary Material}.

The evolutionary structure presented in Figs.~\ref{img:nat-marketsim} and
\ref{img:nat-marketsimintraday} illustrate that the correlation matrix
sometimes maintains its structure for a long time (bright regions),
sometimes changes abruptly (sharp blue stripes), and sometimes returns
to a structure resembling a structure the market has experienced before
(white stripes).  This suggests that the market might move among several
typical market states.  To extract such typical market states, we
perform a clustering analysis in the results of dataset (i).  From our clustering analysis (see {\it
Methods} and {\it Supplementary Material}), we find that there are
``hidden'' states sparsely embedded in time, in addition to regimes that
dominate the market during a continuous period and are easily found
by eye.  For this analysis, we use disjunct two-month time windows
ending at the respective dates. Because of the window length, some
financial crashes cannot be resolved. Our aim is rather to identify the
evolution of the market, which is, in some cases, induced by financial
crisis.  We can confirm in Fig.~\ref{img:nat-typicalstates} that the
typical states obtained from the clustering analysis indeed correspond
to different characteristic correlation structures. To visualize these
characteristic structures, we sort them according to their industry
branch using the Global Industry Classification Standard
(GICS) \cite{gics}. The
industry branches correspond to the blocks on the diagonal.

To visualize the characteristic structures of each state, we calculate its average correlation matrix and sort the companies according to their industry branch, as defined by the Global Industry Classification Standard (GICS). The resulting matrices, the industry branches correspond to the blocks on the diagonal. The correlation between two branches are given by the off-diagonal blocks. The results are illustrated in Fig.~\ref{img:nat-typicalstates}. We can confirm that the
typical states obtained from the clustering analysis indeed correspond
to different characteristic correlation structures.

Our analysis also offers insight into market structure
dynamics. Figure~\ref{img:nat-stateevolution} shows the temporal
behavior of the market state.  The market sometimes remains for a long
time in the same state, and sometimes stays only for a short time.  The
typical duration depends upon the state: Some states (e.g., state 1 and
state 2) appear in clusters in time while other states appear more
sparsely in time (e.g., state 4).  There seems to exist a global trend
on a long time scale, although the market state is switching back and
forth between states.

\begin{figure}[t]
\subfloat[Surface plot]{\includegraphics[width=0.47\textwidth]{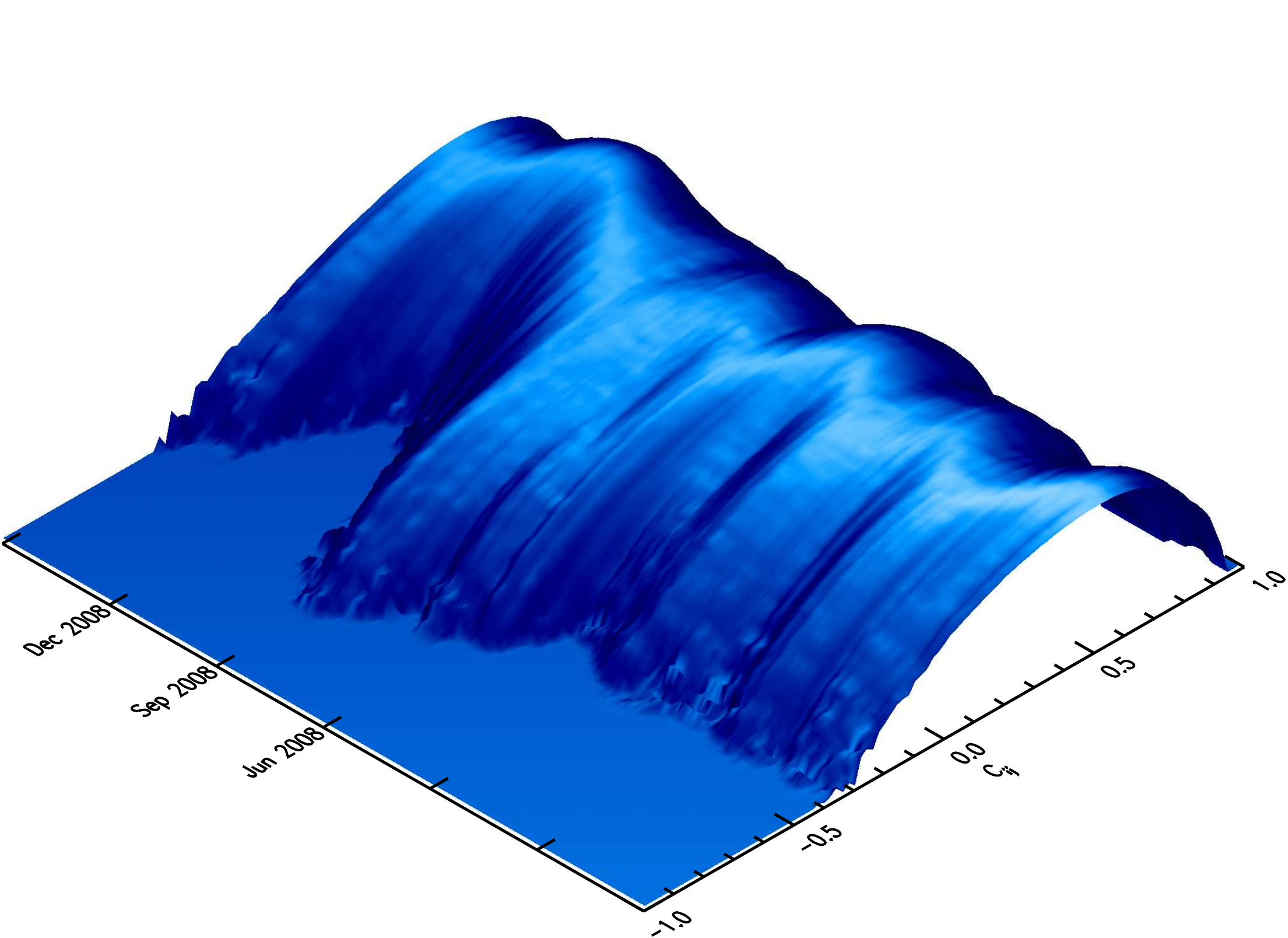}}
\\
\subfloat[Single histograms]{\includegraphics[width=0.47\textwidth]{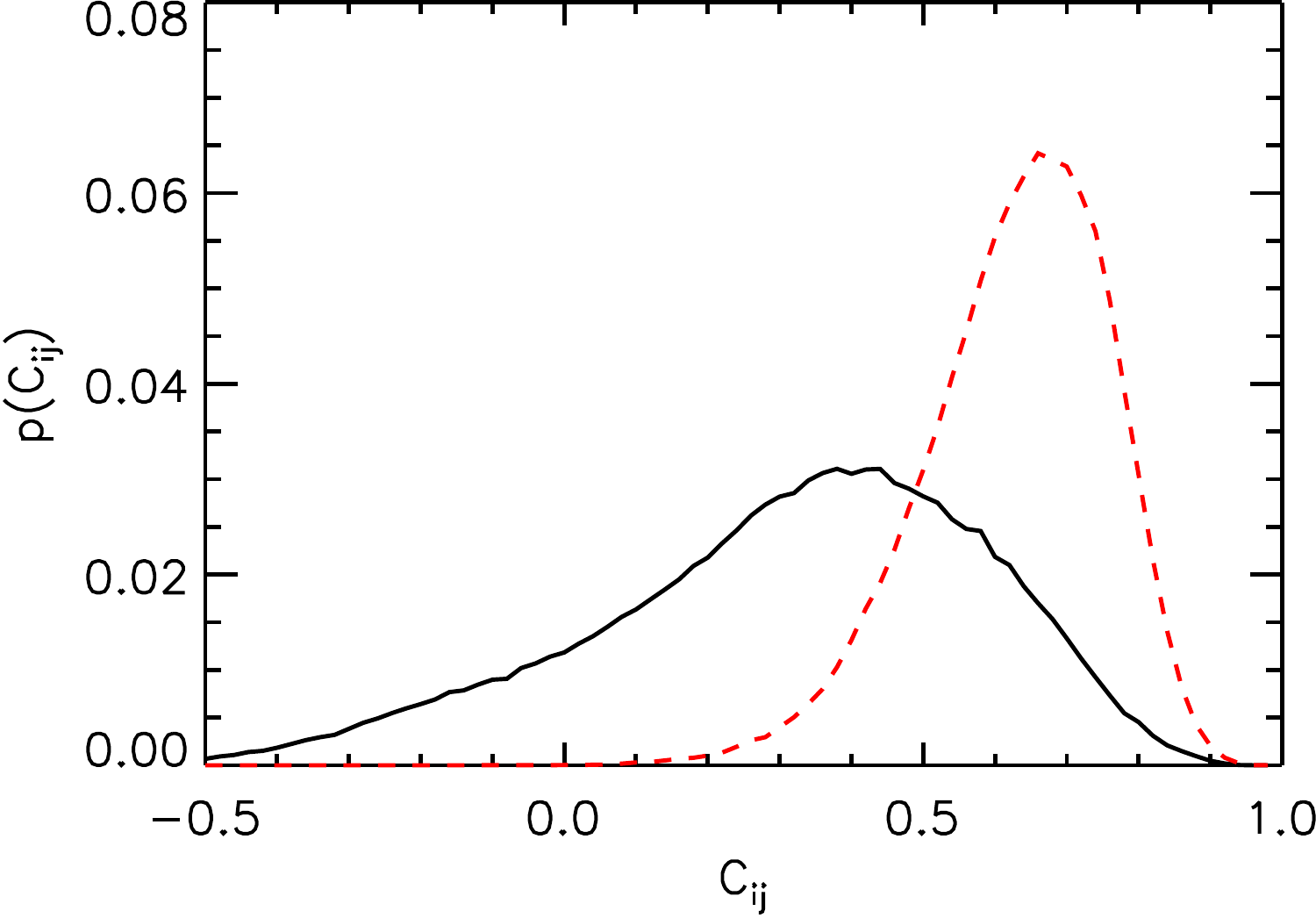}}
\caption{Footprint of the state transition in the 2008 crisis by
  histograms of the correlation coefficients $C_{ij}(t)$. (a) Surface
  plot for the time period September 2007 to March 2009. We use a
  logarithmic scale to show the bimodal structure more clearly. (b)
  Histograms for September 2008 (black solid line) and December 2008
  (red dashed line). }
\label{img:histogram_cij}
\end{figure}

In Fig.~\ref{img:nat-typicalstates}, we can see differences between the
states in the correlation between branches as well as in the correlation
within a branch.  The correlation within the energy, information
technology, and utilities branches is very strong in all states. State 1
shows an overall weak correlation, while states 3 and 4 feature in
addition a strong correlation of the finance branch to other branches.
State 2 shows very unusual behavior: In the period of the dot-com
bubble, many branches are anti-correlated with one another.  In states
5, 6 and 7, the overall correlation level rises, although certain
branches, such as energy, consumer staples, and utilities, are either
strongly or weakly correlated with other branches. The energy branch (E)
can be either strongly correlated to the rest of the market, weakly
correlated, or even anti-correlated.  Therefore we study the histogram
of the correlation coefficients $C_{ij}(t)$. We present the results in
Fig.~\ref{img:histogram_cij}.  In the months leading up to the credit
crunch in October 2008, we observe a bimodal structure in the histogram.
It corresponds to the time period when the Energy branch shows a strong
anti-correlation with other branches. The bimodality suggests that a
subset of stocks -- in this case, predominantly the Energy
stocks -- decouples from the rest of the market.  During the crash, the
histogram shows a very narrow distribution around large values of the
correlation coefficients, which corresponds to state 8 in
Fig.~\ref{img:nat-typicalstates}, where the branch structure is lost
almost completely in an overall strongly correlated market.

\section*{Conclusion}
Our findings offer insight for constructing an ``early warning system'' for financial markets. By providing a simple instrument to identify similarities to previous states during an upcoming crisis, one can judge the current situation properly and be prepared to react if the crisis materializes. Another indication for a crisis is given when the correlation structure undergoes rapid changes.

Using the similarity measure we were able to classify several typical market states between which the market jumps back and forth. Some of these states can easily be identified in the similarity measure. However, there are several states in which the market only stays for a short period. Thus, these states are sparsely embedded in time. With a clustering analysis, we were able to identify these states and disclose a detailed dynamics of the market's state.

A possible application of the similarity measure is risk management.
Given the similarity measure, the portfolio manager is aware of
periods in which the market behaved completely differently and thus can choose not to
include them in his calculations. He can furthermore identify
regions in which the market behaved similarly and refer to these regions
when estimating the correlation matrix. 

Our empirical study is a first step towards the identification of states in financial markets which are a prominent example of complex non-stationary systems.

\section*{Methods}

\subsection{Construction of stock returns}
Let $S$ be the price of a specific stock and $\Delta t$ the interval on
which the return is calculated. For our study, we chose the arithmetic
return, defined as
\begin{equation}
r(t) \equiv \frac{S(t+\Delta t)-S(t)}{S(t)} \ .
\end{equation}
For dataset (i), we chose $\Delta t$ to be 1 day and calculate the stock
returns of each day. For dataset (ii), we chose $\Delta t$ as 1
hour. Furthermore, we obtain this 1-hour return for every minute of a
trading day between 10:45am and 2:45pm.  We obtain the daily data of
dataset (i) from {\it finance.yahoo.com}. The intraday data is obtained from
the New York Stock Exchange's TAQ database.

\subsection{Local normalization}
Sudden changes in drift and volatility can result in erroneous
correlation estimates.  To address this problem, we employ a local
normalization method \cite{schaefer10b}. For each return $r(t)$ we
subtract the local mean and divide by the local standard deviation,
\begin{equation}
\tilde{r}(t)\equiv\frac{r(t)-\left< r(t) \right>_n}{\sqrt{\left< r^2(t)
    \right>_n -\left< r(t) \right>_n^2}} \ . 
\end{equation}
The local average $\left\langle \ldots \right\rangle_n$ runs over the $n$ most recent
sampling points.  For daily data, $n=13$ yields nearly normal
distributed time series, as recently discussed~\cite{schaefer10b}.

\subsection{Outline of top-down clustering}
Our clustering analysis is based on a top-down scheme: All the
correlation matrices are initially regarded as a single cluster and then
divided into two clusters by the procedure based on the k-means
algorithm \cite{macqueen67, faber94, nobuaki04}.  Each division step
consists of the following process:

\begin{enumerate}

\item Choose two initial cluster centers from all matrices. Label all other matrices by the more similar cluster center in terms
  of $\zeta^{(L)}$.
	
\begin{enumerate}
		
\item Recast two new cluster centers to the ``center of mass''
		
\item Re-label all matrices to their most similar cluster center.
		
\item Repeat this process until there is no change in labeling.
	
\end{enumerate}
\end{enumerate}

We stop this division process when the average distance from each 
cluster center to its members becomes smaller than certain threshold. To 
identify the typical market states presented in the manuscript, we chose 
the threshold at 0.1465 as it represents the best ratio between the distances between clusters and their intrinsic radius. One can obtain finer structures by choosing
smaller threshold values, ultimately until all the matrices are 
identified as different components.
The complete results of the clustering analysis are available in the 
{\it Supplementary Material}


\section*{Acknowledgments}
MCM acknowledges financial support from the Fulbright program and from
Studienstiftung des Deutschen Volkes. TS acknowledges support from the JSPS Institutional Program for Young 
Researcher Overseas Visits, Grant-in-Aid for Young Scientists (B) no. 
21740284 MEXT, Japan, and the Aihara Project, the FIRST program from 
JSPS, initiated by CSTP. THS acknowledges support from project 79613 of CONACYT, Mexico.  HES thanks
the NSF for support.

\bigskip

\cleardoublepage
\newpage

\section*{Supplementary Material}
\subsection*{Alternative measure: Difference of largest eigenvalue of correlation matrices}
\begin{figure}[b]
\centering
\includegraphics[width=0.48\textwidth]{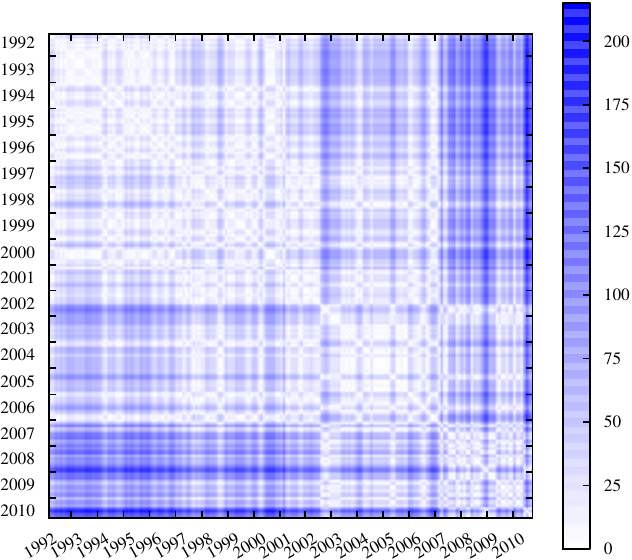}
\caption{Similarly matrix based on the difference of the
  correlation matrices largest eigenvalues.}
 \label{fig:nateigenvaluediff}
\end{figure}

A similar result can be archived using a different approach. The largest
eigenvalue $\lambda_{\mathrm{max}}$ of the correlation matrix $\mathbf
C$ describes the collective motion of all stocks. We can also define the
similarity measure by the distance of these eigenvalues,
\begin{equation}
\zeta_{\mathrm{alt}}(t_{1},t_{2})\equiv \left| \lambda_{\mathrm{max}}(\mathbf C(t_{1}))
- \lambda_{\mathrm{max}}(\mathbf C(t_{2})) \right| \ .
\label{eq:nat-simmeasure}
\end{equation}

Figure~\ref{fig:nateigenvaluediff} illustrates that this leads to an
almost identical result. The advantage of this technique is that the
noise in the correlation matrix only contributes to small eigenvalues \cite{laloux99, plerou03}. Thus, by only taking
into account the largest one, we filter out the noise.  However, this
approach also presumes that the corresponding eigenvector does not
change. Our results indicate that the largest eigenvalue almost remains
constant, but this might not always be the case. Especially during financial
crises.

\subsection*{Stable period within 2008-2009 crisis}

\begin{figure}[b]
\centering
\subfloat[Crisis (2008/10/15 - 2009/4/1, excluding stable period)]{\includegraphics[width=.4\textwidth]{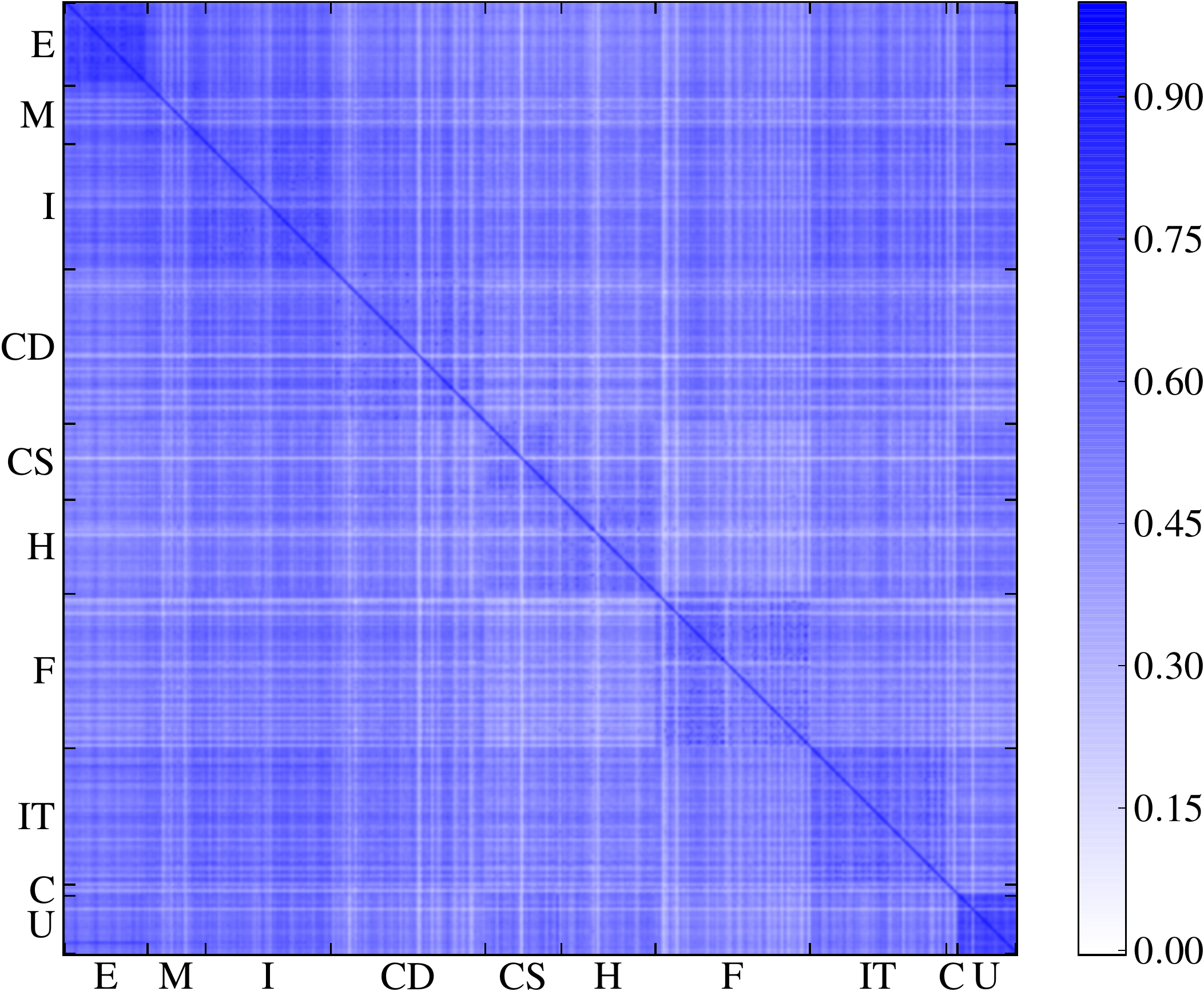}}
\\
\subfloat[Stable period (2009/1/1 - 2009/1/21)]{\includegraphics[width=.4\textwidth]{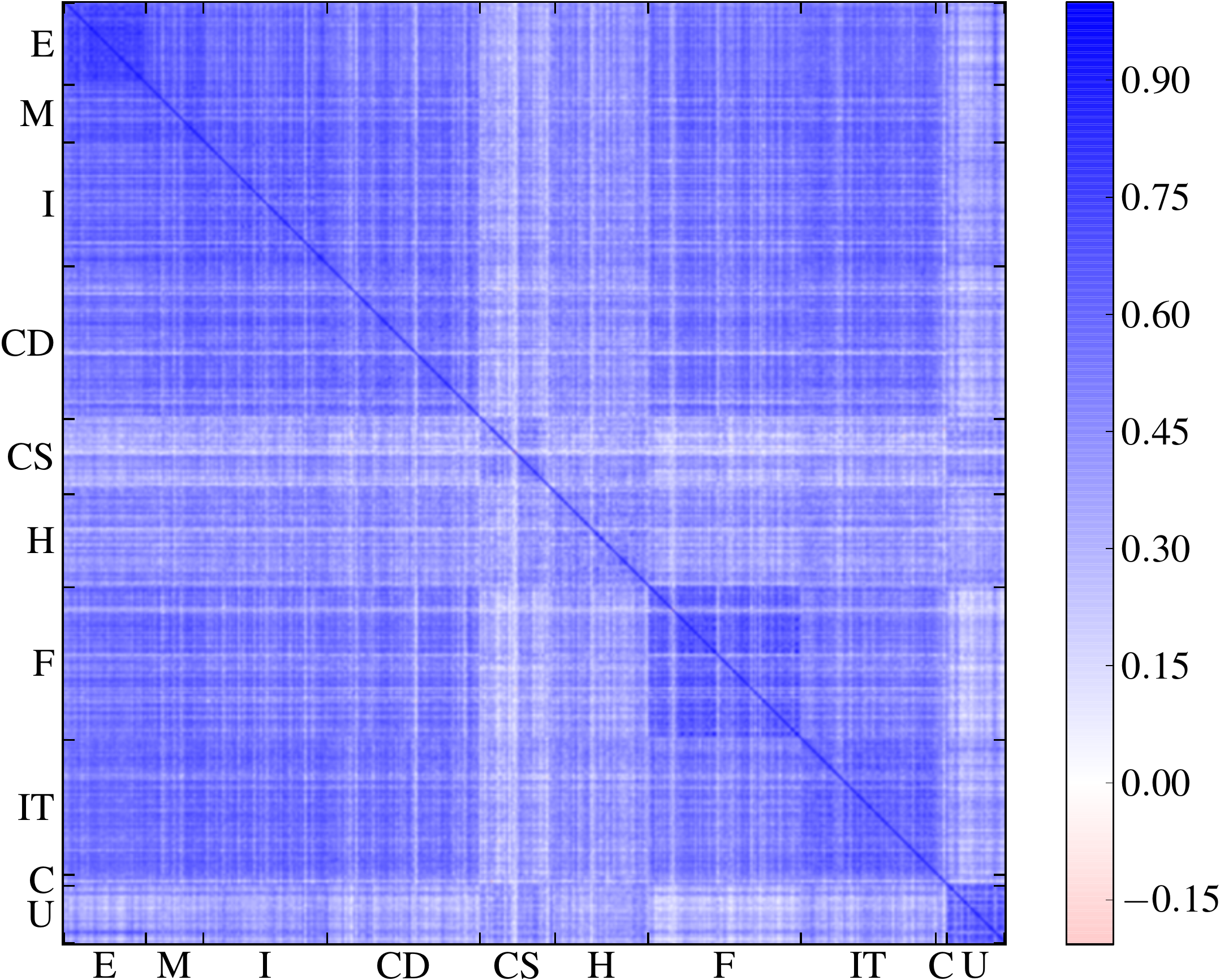}}
\caption{Within the 2008--2009 crisis, the market temporarily
  stabilizes. This stable state is very similar to the pre-crisis state
  that we identified from daily data (state 7).}
\label{img:nat-stable-within-crisis}
\end{figure}

A detailed look of the correlation structure of the 2008-2009 crisis can be found in
Fig.~\ref{img:nat-stable-within-crisis}. While during the crisis, an
overall high correlation level dominates, the structure stabilizes for a
short time of 3 weeks. During this stable period, the structure is very
similar to state 7, that occurred just just before the crisis.

\subsection*{Difference matrices to average correlation matrix}

\begin{figure*}[pb]
\centering
\subfloat[state 1]{\includegraphics[width=0.23\textwidth]{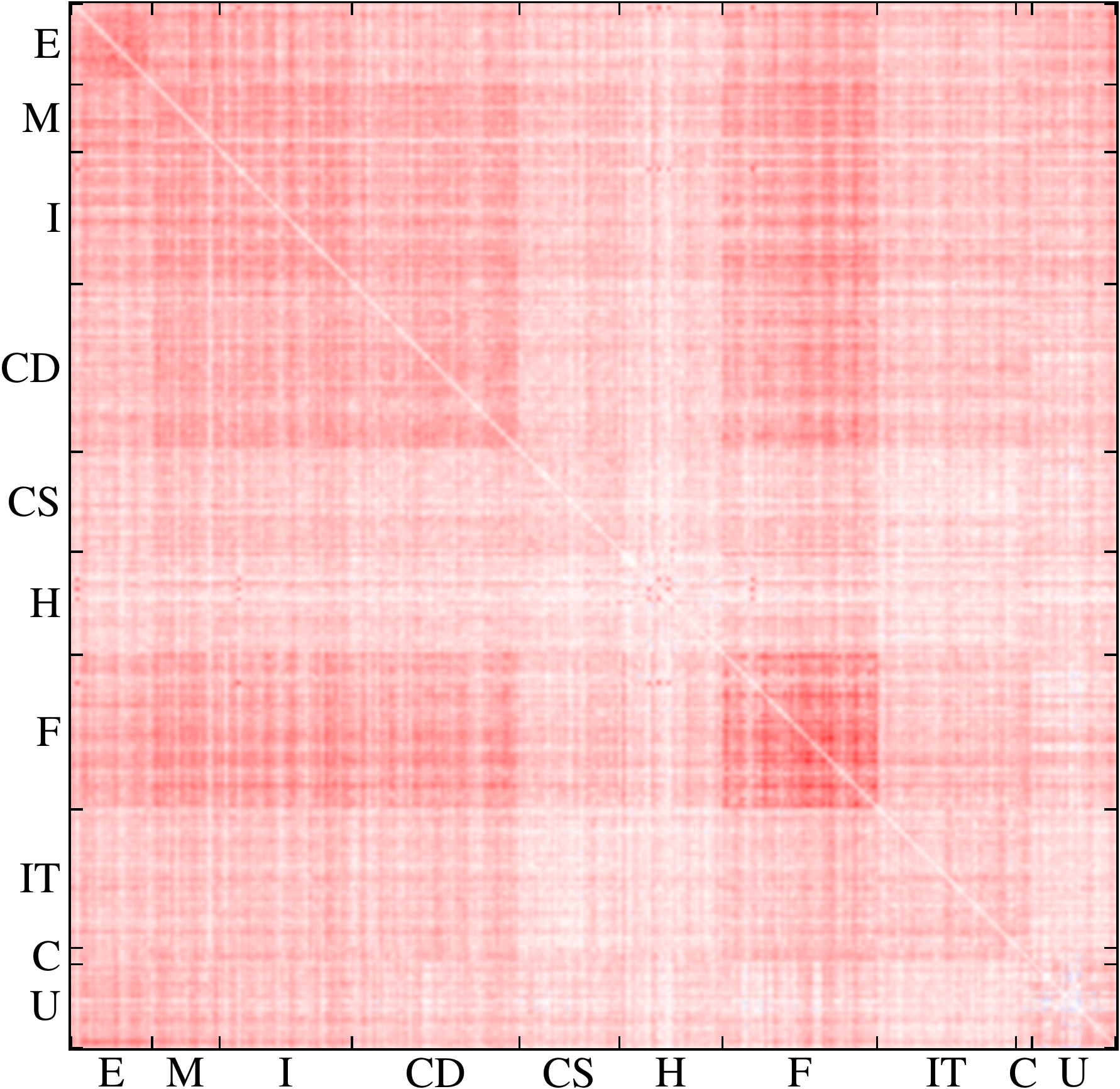}}
\quad
\subfloat[state 2]{\includegraphics[width=0.23\textwidth]{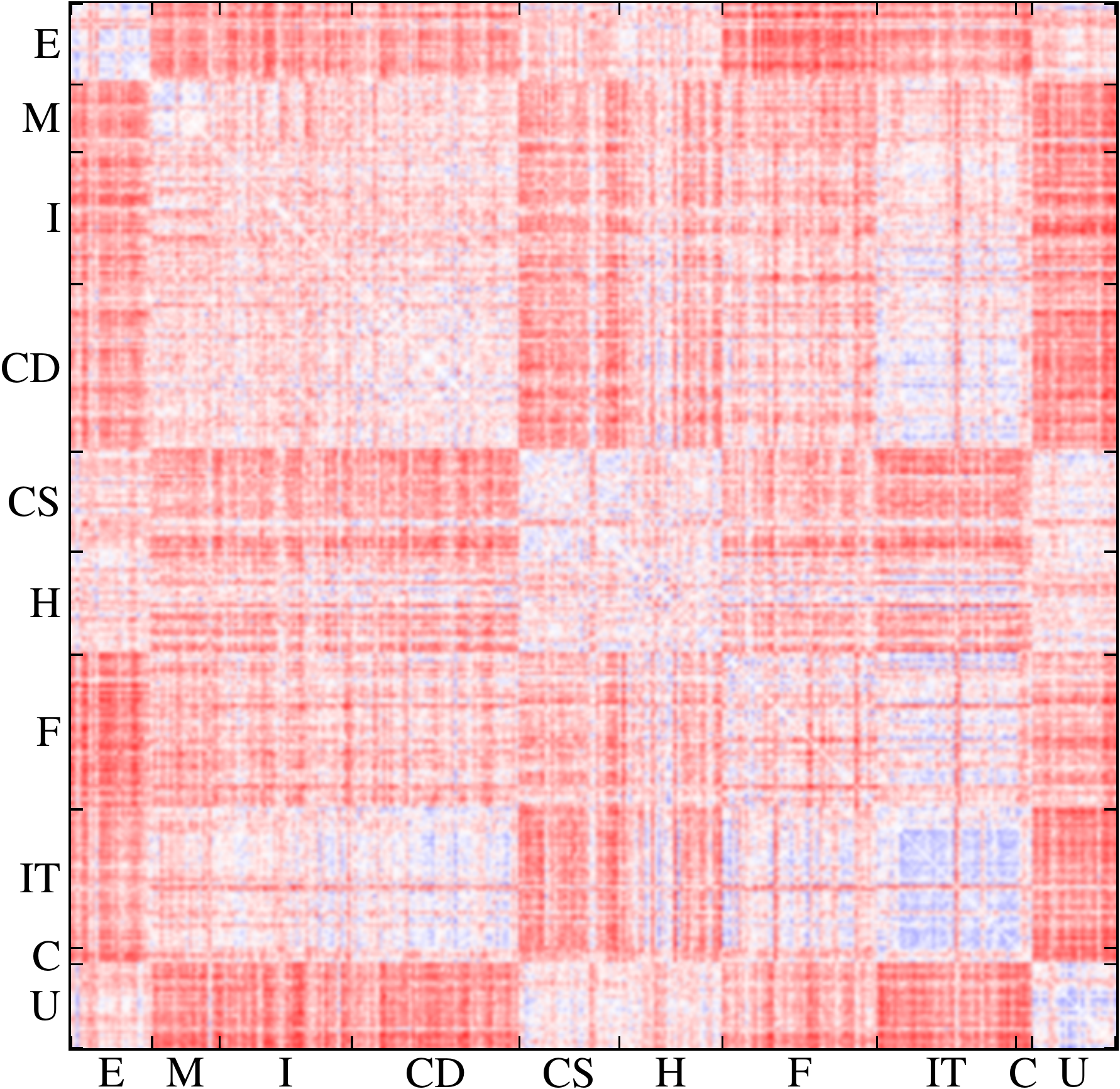}}
\quad
\subfloat[state 3]{\includegraphics[width=0.23\textwidth]{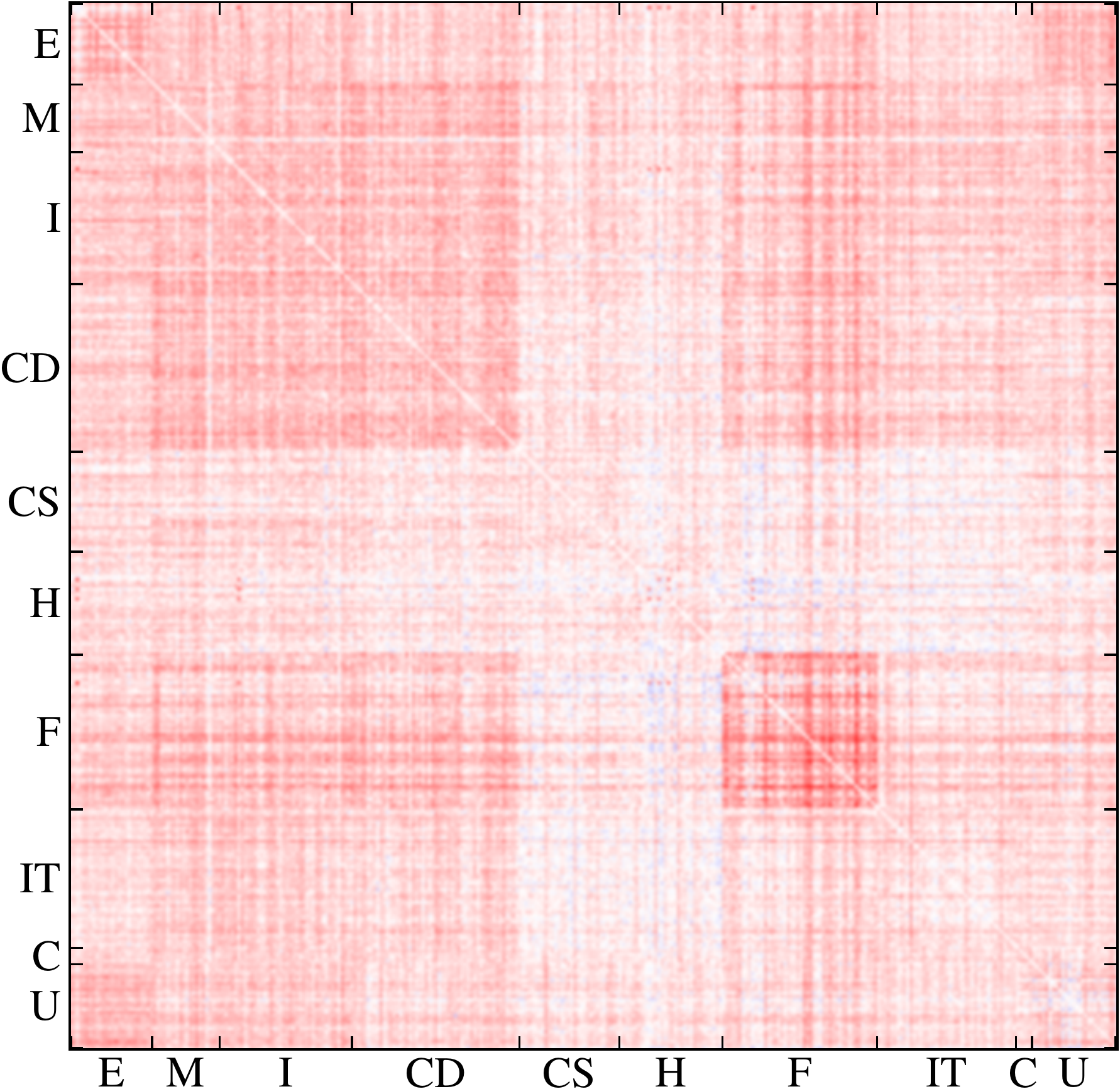}}
\quad
\subfloat[state 4]{\includegraphics[width=0.23\textwidth]{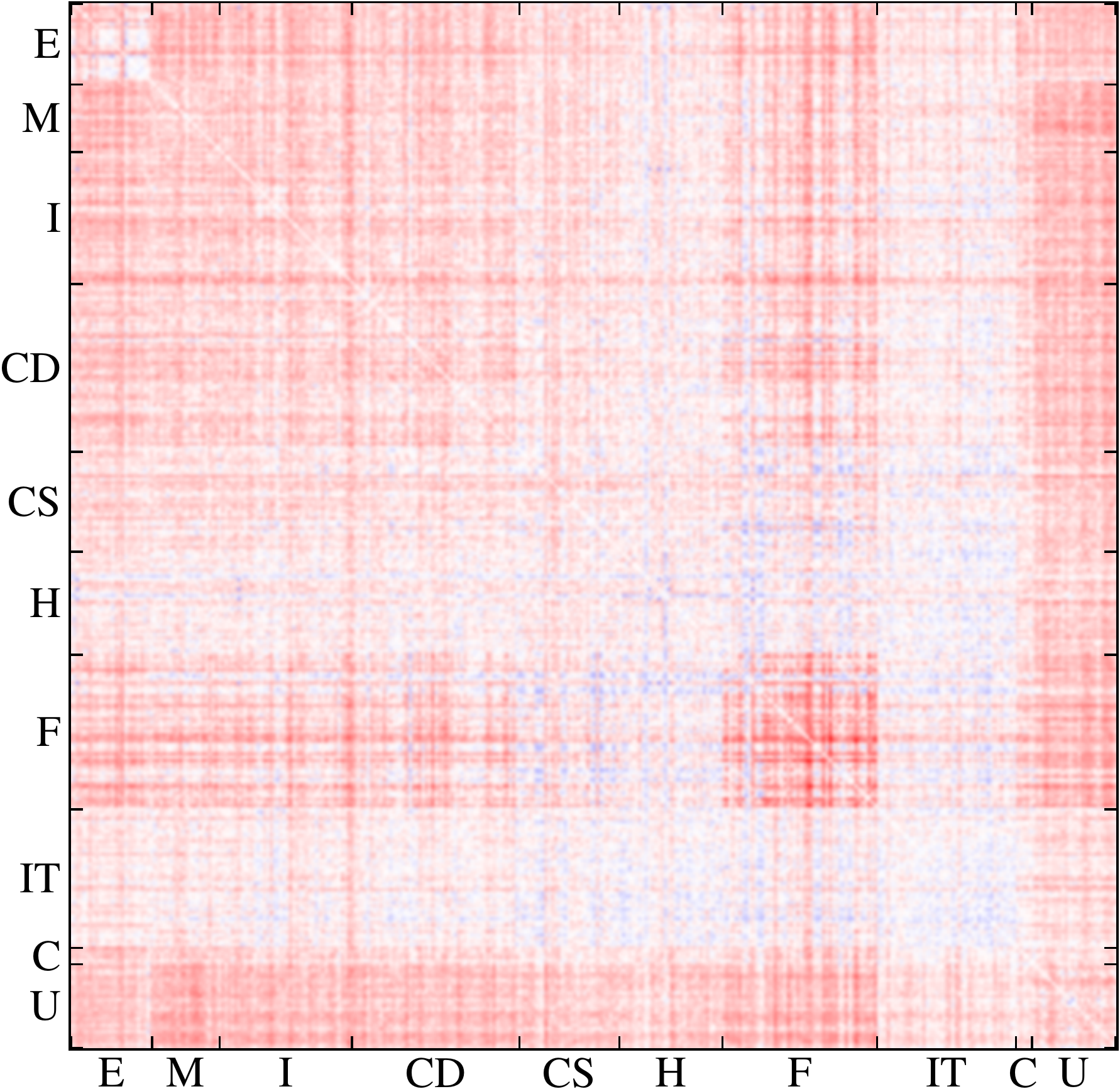}}
\\
\subfloat[state 5]{\includegraphics[width=0.23\textwidth]{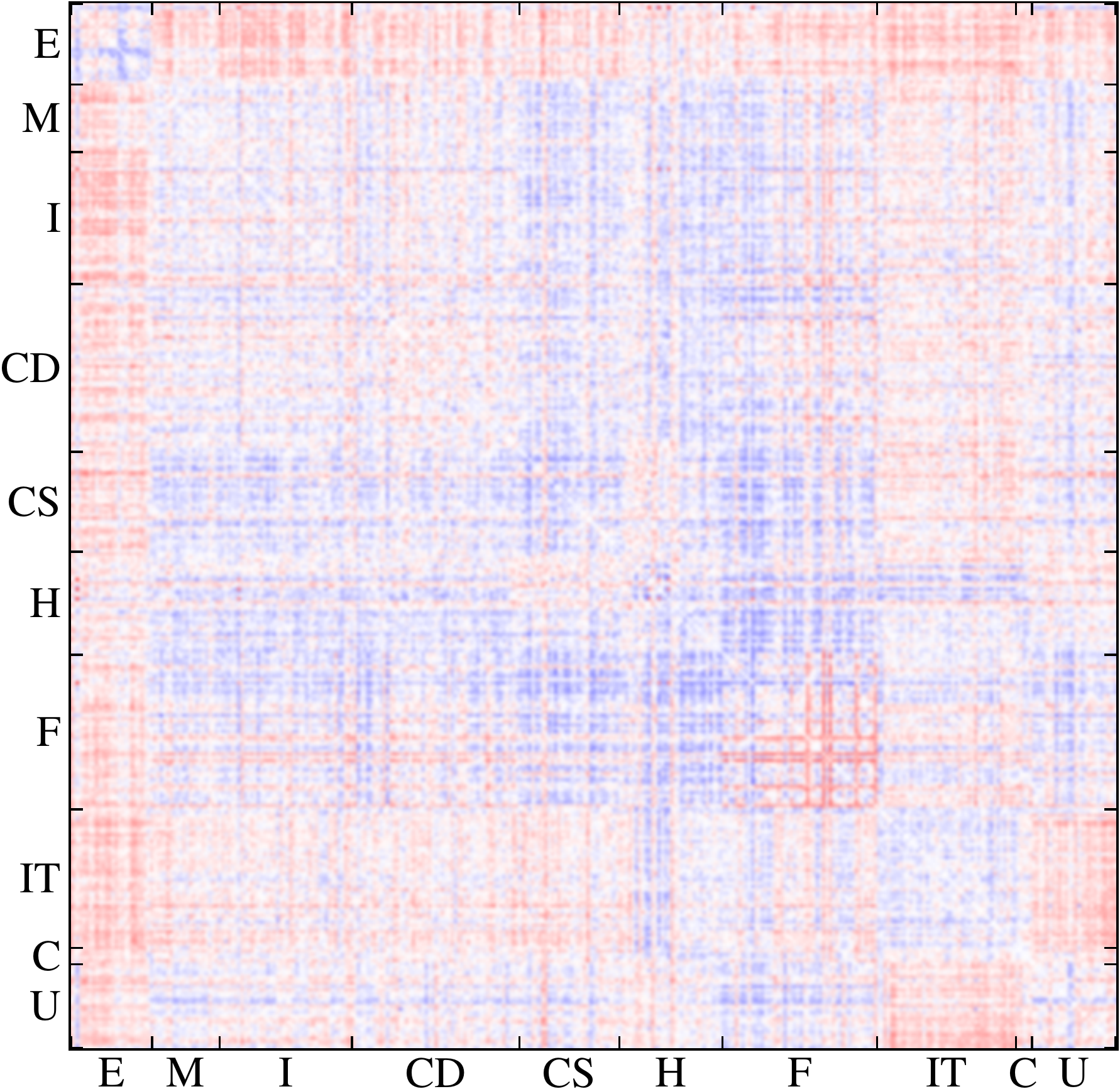}}
\quad
\subfloat[state 6]{\includegraphics[width=0.23\textwidth]{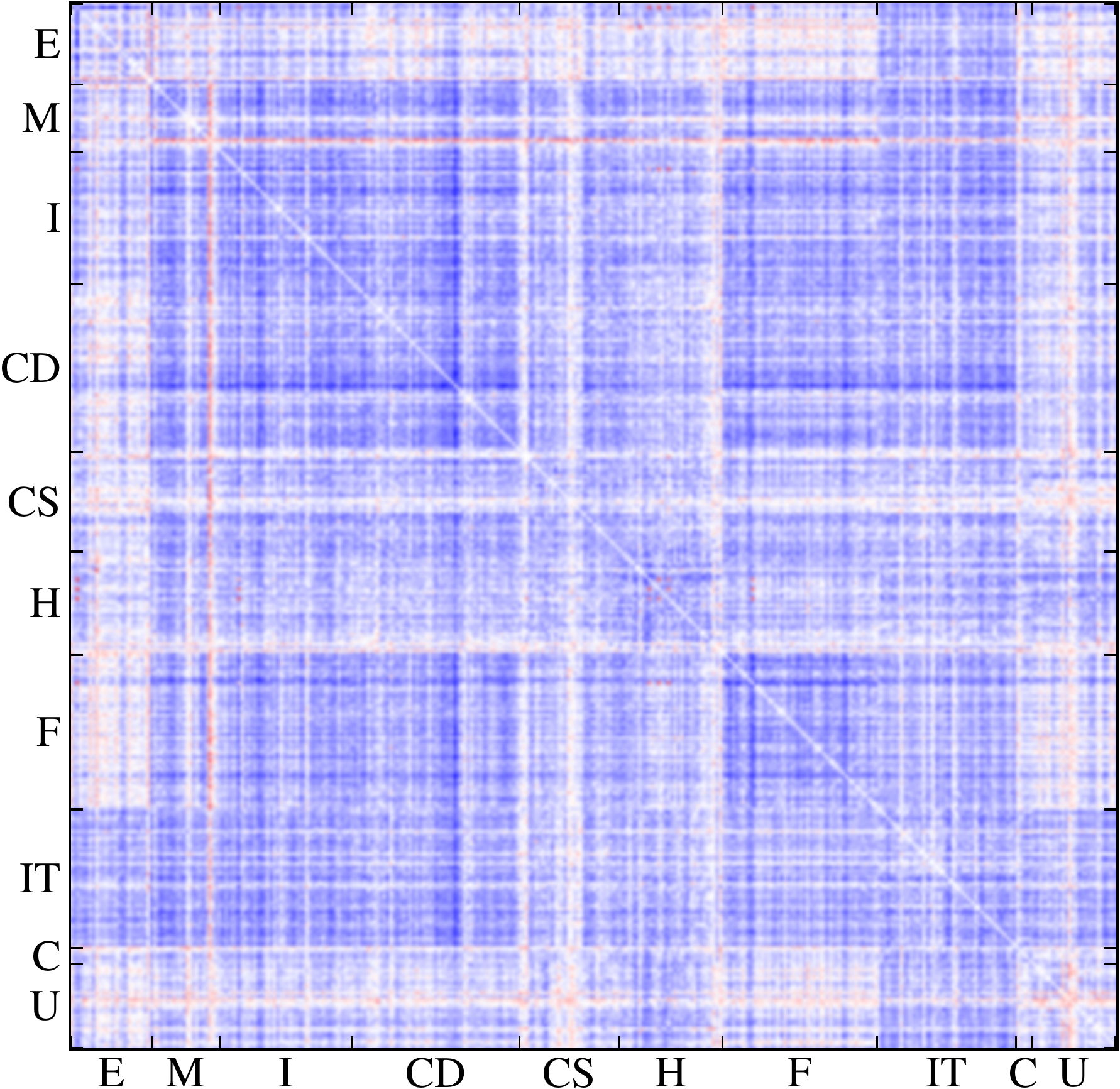}}
\quad
\subfloat[state 7]{\includegraphics[width=0.23\textwidth]{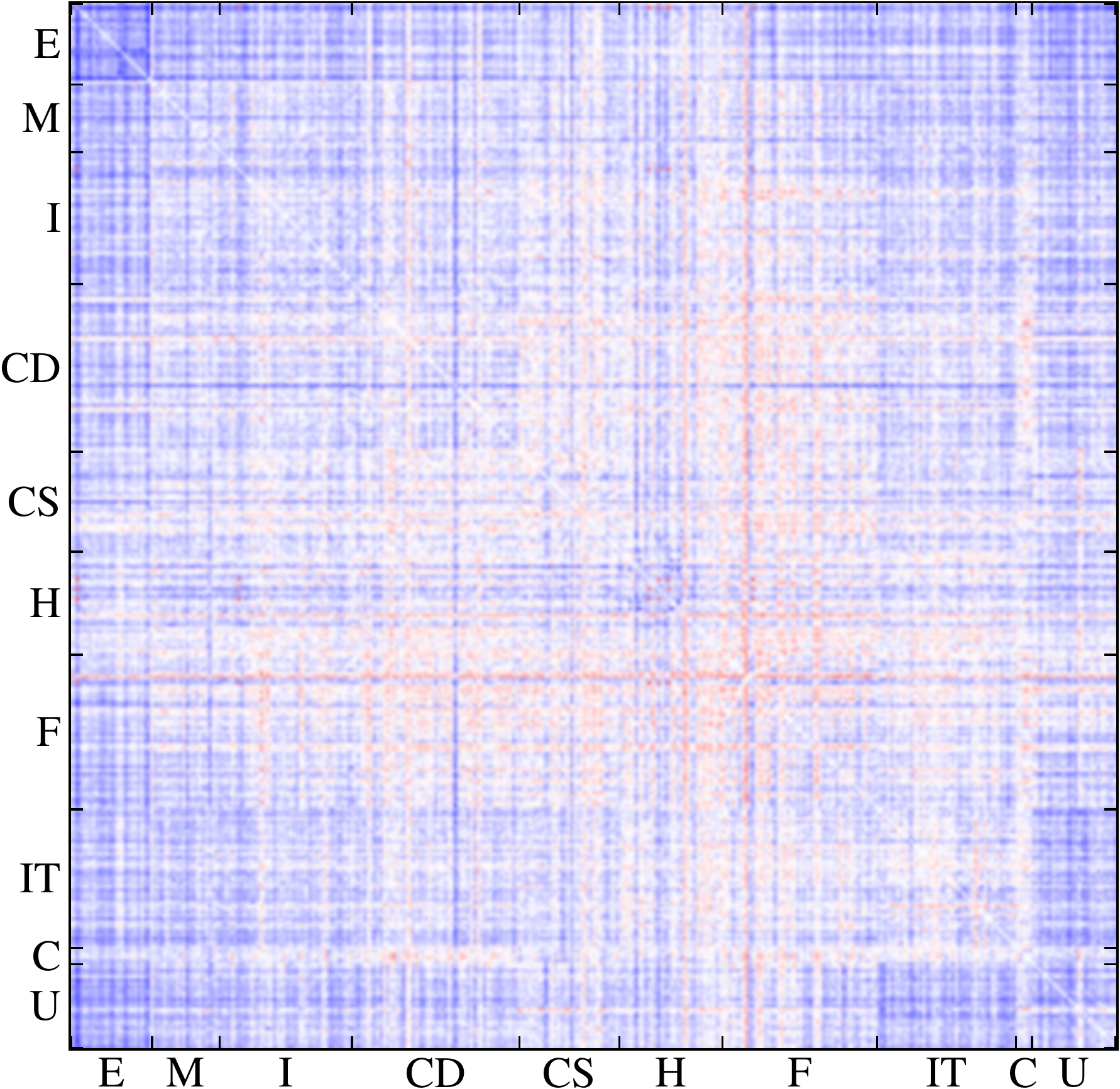}}
\quad
\subfloat[state 8]{\includegraphics[width=0.23\textwidth]{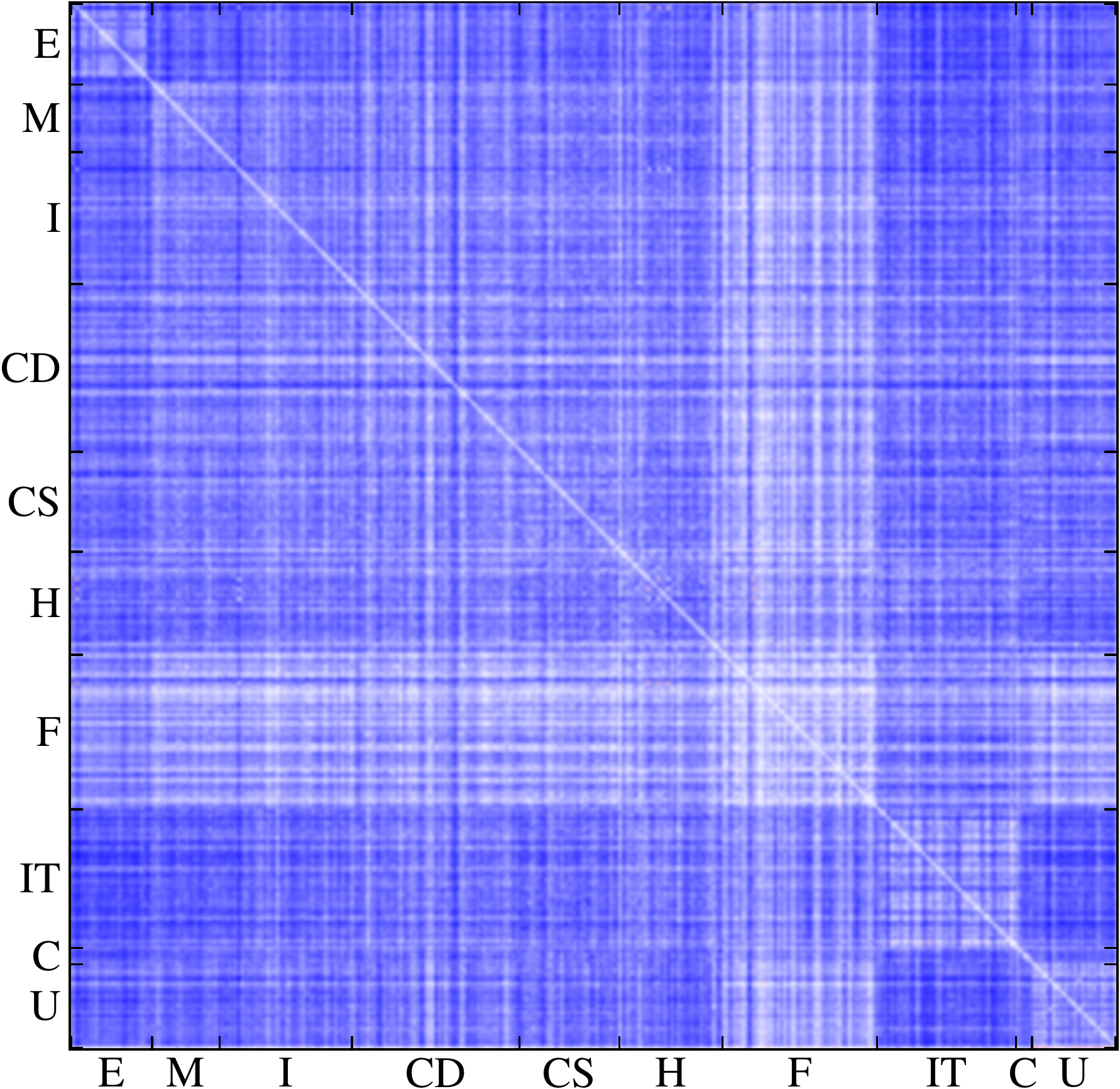}}
\\
\subfloat[Overall average
  correlation]{\includegraphics[width=0.45\textwidth]{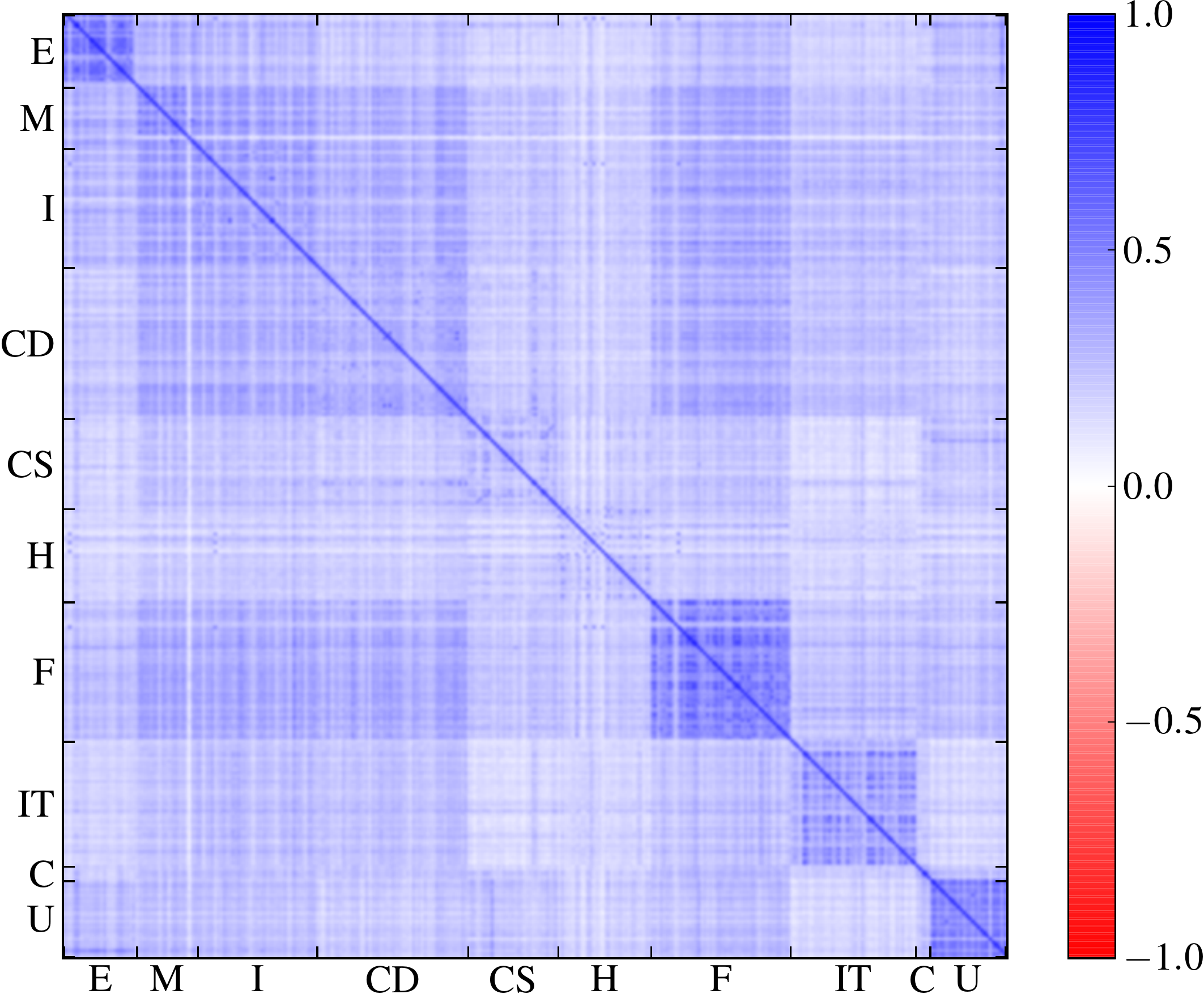}} 
\caption{(a)--(h): Difference of the states' correlation matrices to
  average correlation matrix (i).}
 \label{fig:matdifference}
\end{figure*}

Some of the correlation structures in Fig.~\ref{img:nat-typicalstates}
look quite similar at first sight. Their distinctiveness can be emphasized by
calculating the difference to the average correlation level. This is
shown in Fig.~\ref{fig:matdifference}. For example, state 3 and 4 look
very similar in Fig.~\ref{img:nat-typicalstates}. However,
Fig.~\ref{fig:matdifference} unveils that the correlation within the
Energy sector (E) is completely diverse.

\clearpage

\vfill

\begin{figure*}[p]
\centering
\includegraphics[width=0.65\textwidth]{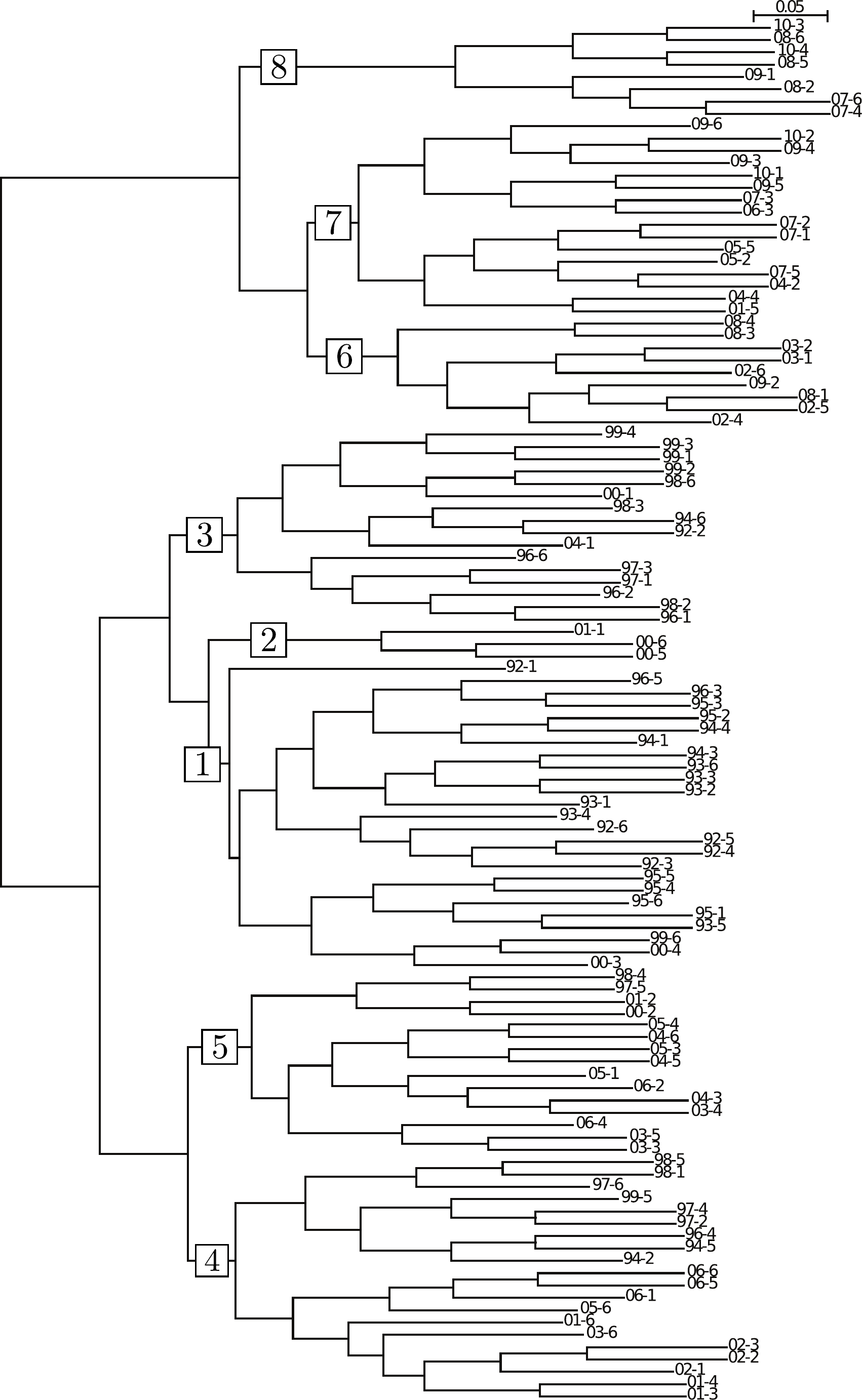}
\caption{The entire tree of the clustering analysis presented here for
  the threshold $= 0$: No termination of the division process takes
  place until all the correlation matrices are identified as different
  components. The large bold numbers represent the market states each of
  which consists of the matrices in the sub-trees below. Each right end
  of the tree corresponds to each 2-month term (year-term). Terms 1, 2,
  \dots, 6 correspond to January to February, March to April, \dots,
  November to December, respectively. The length of each branch
  represents the distance from the center of the subcluster to the
  center of the original cluster before the last dual division.}
\label{img:nat-topdown-complete}
\end{figure*}

\end{document}